\documentclass[aps,prf,reprint,superscriptaddress]{revtex4-1}
\usepackage[utf8]{inputenc}
\usepackage{bm}
\usepackage{caption}
\usepackage{fullpage}
\usepackage{amsmath}
\usepackage{amssymb}
\usepackage{amsbsy}
\usepackage{graphicx}
\usepackage{indentfirst}
\usepackage{setspace}
\usepackage{color}
\usepackage{natbib}
\usepackage{hyperref}
\usepackage{cleveref}
\usepackage{verbatim}
\usepackage{subcaption}
\usepackage{graphicx}
\usepackage{xcolor}
\usepackage[b]{esvect}
\usepackage{gensymb}

\newcommand\etal{\mbox{\textit{et al.}}}
\newcommand{\grad}{{\mbox{\boldmath $\nabla$}}}
\newcommand{\bdot}{{\mbox{\boldmath $\cdot$}}}
\newcommand{\bu}{{\bf u}}

\newcommand{\btimes}{{\mbox{\boldmath $\times$}}}
\newcommand{\bB}{{\bf B}}

\begin{document}

\title{Effects of External Magnetic Fields on the Multi-mode Rayleigh-Taylor Instability}

\author{Xin Bian}
\affiliation{Department of Mechanical Engineering, University of Rochester, Rochester, NY 14627, USA}
\author{Riccardo Betti}
\affiliation{Department of Mechanical Engineering, University of Rochester, Rochester, NY 14627, USA}
\affiliation{Laboratory for Laser Energetics, University of Rochester, Rochester, NY 14623, USA}
\affiliation{Department of Physics and Astronomy, University of Rochester, Rochester, NY 14627, USA}
\author{Dongxiao Zhao}
\email{zhaodongxiao1990@gmail.com}
\affiliation{Department of Mechanical Engineering, University of Rochester, Rochester, NY 14627, USA}
%\affiliation{Laboratory for Laser Energetics, University of Rochester, Rochester, NY 14623, USA}
\author{Hussein Aluie}
\affiliation{Department of Mechanical Engineering, University of Rochester, Rochester, NY 14627, USA}
\affiliation{Laboratory for Laser Energetics, University of Rochester, Rochester, NY 14623, USA}
\affiliation{Department of Mathematics, University of Rochester, Rochester, New York }

\begin{abstract}
The magneto–Rayleigh–Taylor instability (mRTI) is a key process in inertial confinement fusion and is thought to be widespread in the interstellar medium, where it can concentrate plasma into discrete structures. We present resistive MHD simulations of the nonlinear evolution of multi‑mode mRTI in both two and three dimensions, examining the effects of uniform external magnetic fields oriented either parallel or perpendicular to the initial interface. In both 2‑D and 3‑D, weak parallel fields enhance mixing‑zone growth, whereas stronger fields suppress it. For perpendicular fields, growth is initially inhibited but becomes enhanced at later times. These behaviors arise from magnetic tension, which modifies flow anisotropy, buoyancy, drag, and vortex dynamics. The interplay of these mechanisms governs the distinct ways in which magnetic fields influence mRTI evolution.

\end{abstract}
\maketitle
\section{Introduction}

Decades of research have been devoted to the Rayleigh–Taylor instability (RTI), which arises when a light fluid of density $\rho_l$ accelerates into a heavy fluid of density $\rho_h$, or equivalently when a heavy fluid is supported against gravity by a lighter one \cite{rayleigh1883analytic,taylor1950instability,sharp1984overview,abarzhi2010review,boffetta2017incompressible,ZHOU20171,zhou2017rayleigh,ZhouARFM2025}. In the incompressible, inviscid, and immiscible limit, and for perturbation wavelengths much smaller than the domain size, high–wavenumber modes grow faster during the linear stage, with a growth rate $\gamma = \sqrt{g k A}$, where $k$ is the wavenumber and $A \equiv (\rho_h - \rho_l)/(\rho_h + \rho_l)$ is the Atwood number. As the perturbation amplitude becomes comparable to its wavelength, RTI enters the nonlinear stage, in which coherent bubble and spike structures emerge, driven in part by the Kelvin–Helmholtz instability (KHI). In the fully developed nonlinear regime, self‑similar theory \cite{dimonte2004comparative,Zhouetal2021PhysD,Zhou2024Book} predicts that the bubble height $h_b$ grows as  
\begin{eqnarray}
\label{selfsimilar}
h_b \propto \alpha_b A g t^2,
\end{eqnarray}  
where $\alpha_b$ is a dimensionless constant, $g$ is the gravitational acceleration, and $t$ is time.

RTI is prevalent in a wide range of astrophysical environments, including the evolution of supernova explosions \cite{arnett1989supernova}, young supernova remnants \cite{jun1996interaction,blondin2001rayleigh}, cold fronts in the intergalactic medium \cite{vikhlinin2002cold}, and accretion disks \cite{krumholz2009formation}. In inertial confinement fusion (ICF), RTI is a major obstacle to achieving ignition, as instability–induced mixing between the hotspot and the surrounding cold deuterium–tritium fuel significantly degrades hotspot performance \cite{betti2016inertial,woo2018effects}. When magnetic fields are present, RTI couples to the field to form the magneto–Rayleigh–Taylor instability (mRTI), which plays a critical role in magnetized liner inertial fusion (MagLIF) \cite{slutz2012high} and in many astrophysical plasmas. Examples include filamentary structures in the solar corona \cite{isobe2005filamentary}, the clumpy radio shell morphology of Type‐I supernova remnants \cite{jun1995mhd}, and the emission‐line filaments in the Crab Nebula \cite{hester1996wfpc2,hester2008crab}.  

\subsection{Linear stability analysis}
The studies on mRTI are relatively few compared to hydrodynamic RTI (hRTI). Chandrasekhar \cite{chandrasekhar1961hydrodynamic} conducted the linear stability analysis of stratified, incompressible, inviscid, non-resistive mRTI. 
For a uniform magnetic field $\bB$ perpendicular to the initial perturbation plane, the analysis suggests that the growth rate is reduced compared to the hydrodynamic case, and the growth rate saturates at high wavenumber dependent on the magnetic field strength. The growth rate satisfies \cite{jun1995numerical,Briard2024JFM}
\begin{eqnarray}
\begin{split}
    \gamma^3 + \gamma^2 k V_A (\sqrt{\alpha_1} + \sqrt{\alpha_2})+ (k^2V_A^2-Agk)\gamma \\= \frac{2Ag V_Ak^2}{\sqrt{\alpha_1} + \sqrt{\alpha_2}}
\end{split} \label{eq:vB_growth}
\end{eqnarray}
where $\alpha_1 = \rho_l/\rho_0, \alpha_2 = \rho_h/\rho_0, \rho_0 = (\rho_l+\rho_h)/2$, and $V_A=B/\sqrt{\rho_0}$ is the Alfven velocity \footnote{We adopt the non-dimensional unit here, the factor of $1/\sqrt{4\pi}$ is absorbed into $\bB$.}. 
According to Eq.~(\ref{eq:vB_growth}), the mRTI growth rate is lower than that of the hRTI. However, in this configuration, there is no cutoff wavenumber, and at large $k$ the growth rate saturates at $\gamma=2Ag/[V_A(\sqrt{\alpha_1}+\sqrt{\alpha_2})]$.

For a magnetic field $\mathbf{B}$ oriented parallel to the initial perturbation plane, linear stability analysis shows that interchange modes (wavevectors perpendicular to $\mathbf{B}$) remain unaffected, whereas undular modes (wavevectors parallel to $\mathbf{B}$) are stabilized by the field. The corresponding growth rate is  
\begin{eqnarray}
\gamma = \sqrt{g k A - \frac{2 B^2 k^2 \cos^2\theta}{\rho_h + \rho_l}},  
\label{eq:lsa}
\end{eqnarray}  
where $\theta$ is the angle between the perturbation wavevector $\mathbf{k}$ and the magnetic field $\mathbf{B}$ \cite{chandrasekhar1961hydrodynamic}. Equation~(\ref{eq:lsa}) indicates that a horizontal magnetic field, similar to the vertical field case, also reduces the growth rate relative to the purely hydrodynamic RTI. However, unlike the vertical–field case which approaches a finite asymptotic growth rate at large $k$, the parallel–field configuration exhibits a finite cutoff wavenumber beyond which all modes are completely stabilized. Equivalently, the critical magnetic field strength $B_c$ required to suppress an undular mode of wavelength $\lambda = 2\pi/k_c = L_x$, where $L_x$ is the horizontal domain size, is  
\begin{eqnarray} \label{eq:critical_B}
B_c = \sqrt{\frac{(\rho_h - \rho_l) g \lambda}{4\pi}} .
\end{eqnarray}

In contrast to Chandrasekhar's linear stability analysis \cite{chandrasekhar1961hydrodynamic} presented above, recent analytical and numerical studies \cite{gupta2010effect,khan2011development} on the early nonlinear regime suggested that the parallel magnetic field might stabilize the interchange modes, while in the self-similar turbulent stage \cite{Briard2024JFM,SinghPal2025JFM} vertical magnetic fields could enhance the mixing zone growth in a Boussinesq MHD setting.

In addition, the effects of additional physics such as compressibility, ablation, stratification, and rotation on mRTI were also studied analytically and numerically \cite{liberatore2008analytical, duan2018magneto,ZhangYan2022PoP,ZhangYan2024JFM}, with varying degrees of influence on mRTI mixing zone growth. In the follows, we shall review previous works regarding the magnetic field effects on the mixing width development, including both 2-D and 3-D results.

\subsection{2D mRTI}
2D mRTI was studied using numerical simulations \cite{jun1995numerical,chambers2011magnetic,perkins2013two, Srinivasan2012,srinivasan2013mitigating,perkins2017potential}. Jun \etal \cite{jun1995numerical} showed that both parallel and vertical (to the perturbation plane) fields stabilize single-mode mRTI, and parallel fields are more effective at the stabilization. For multi-mode mRTI, parallel fields have stabilization effects, while vertical fields might enhance or suppress the growth, depending on the field strength. However, the resolutions and aspect ratio of the computational grids ($120\times80$) are limited by today's standard. Higher resolutions and larger domains are required to investigate 2D nonlinear development. 
Srinivasan and Tang \cite{srinivasan2013mitigating} suggested applying a strong external magnetic field to suppress mRTI and reduce mixing using ICF relevant parameters. 2D simulations on National Ignition Facility (NIF) targets showed that an initial axial magnetic field can suppress RTI \cite{perkins2013two,perkins2017potential}. In addition to an external magnetic field, self-generated magnetic fields could reach $1.6\times10^3-6\times10^3$ Tesla in a NIF target \cite{Srinivasan2012, srinivasan2013mitigating,garcia2020self,garcia2021magnetic,ZhangYan2022PoP,garcia2022theory}. The results show that the self-generated magnetic field is too weak to affect RTI growth in ICF directly, but the thermal energy loss might be reduced significantly ($\approx$ 2-10) via inhibiting the electron heat conduction. 

Even though the above 2D studies \cite{Srinivasan2012, srinivasan2013mitigating, perkins2013two, perkins2017potential} suggested promising results by applying magnetic fields in ICF, we note that mRTI could be very different in 2D and 3D. 
For example, the results in Refs. \cite{tryggvason1990computations,youngs1991three,Cabot06POF, Zhao2025JFM} show that the growth rates of hRTI are different between 2D and 3D. Ablative RTI can be destabilized more easily in 3D than in 2D \cite{zhang2018self,xin2019two,zhang2020nonlinear}. 3D mRTI is still unstable even with a strong magnetic field ($|\bB|>B_c$) parallel to the perturbation plane due to the presence of two independent interchange modes, whereas 2D mRTI is fully suppressed.

\subsection{3D parallel magnetic fields}
For parallel external magnetic fields (aligned with the initial interface) in 3D mRTI, existing studies of the nonlinear regime are relatively scattered. Stone and Gardiner \cite{stone2007magnetic,stone2007nonlinear} investigated the 3D nonlinear evolution at $A = 0.5$ and $0.82$ in a domain with a horizontal-to-vertical aspect ratio of $1:1:2$, considering external magnetic field strengths from $0$ to $0.6B_c$ ($B_c$ the critical field strength). They found that parallel magnetic fields of varying strengths enhance mRTI development compared to the hydrodynamic case, primarily due to the mixing suppression by magnetic fields.  
  
Using a different configuration with an aspect ratio of $1:0.25:4$ and a density ratio of $10:1$, Carlyle and Hillier \cite{carlyle2017non} reported that the bubble growth rate generally decreases as the magnetic field strength increases from $0.6B_c$ to $1.3B_c$, a range larger than that considered by Ref.~\cite{stone2007nonlinear}. However, the suppression was non-monotonic with increasing field strength. They attributed the overall reduction to magnetic tension in stronger fields. Their simulations also revealed asymmetric behavior between bubbles and spikes. Notably, the narrow $y$-extent ($1:0.25:4$) allowed for only limited number of bubbles across the spanwise direction, constraining the development of interchange modes, which is known to be influenced by the magnetic field from the onset of the nonlinear stage \cite{gupta2010effect,khan2011development}. Such domain width limitations may also inhibit the onset of self-similar growth in the late stage \cite{dimonte2004comparative}.  
  
Both Refs.~\cite{stone2007nonlinear} and \cite{carlyle2017non} have adopted an ideal compressible MHD equations that dampens small-scale structures by numerical viscosity and numerical magnetic diffusivity, whose effects might be case dependent. In contrast, a recent study by Kalluri and Hillier \cite{Kalluri2025JFM} employed a variable-density incompressible viscous MHD model with a density ratio of $3:1$ and relatively low magnetic fields ($0$ to $0.25B_c$). They found that the mixing width growth rate increases with stronger parallel magnetic fields, complementing the ideal MHD results of \cite{carlyle2017non} at higher density ratio.  A similar conclusion was drawn in \cite{BriardGrea2025} for horizontal magnetic fields ranging from $0$ to approximately $0.42B_c$, where it was observed that the growth of the mixing width is nearly doubled at the largest investigated external magnetic field. Furthermore, an analytical relation linking the growth rate, mixing, anisotropy, and induced magnetic fields was derived.
  
Overall, the differing trends reported across these studies highlight the need for further investigation to resolve these discrepancies, explicitly examining the roles of physical viscosity and magnetic diffusivity under parameters relevant to realistic flow scenarios.

\subsection{3D vertical magnetic fields}

In addition to parallel magnetic fields, mRTI with vertical magnetic fields is significant in astrophysical flows and ICF. For instance, white dwarfs can generate strong magnetic fields with strengths reaching $10^8$ to $10^9$ Gauss \cite{ghezzi2001magnetic}. During a Type Ia supernova explosion in a white dwarf, plasmas near the equatorial region encounter magnetic fields perpendicular to the direction of gravity, corresponding to the parallel magnetic field case. Conversely, plasmas near the magnetic poles are influenced by magnetic fields aligned with the direction of gravity, representing the vertical magnetic field case \cite{gupta2010effect}. Previous two-dimensional cylindrical studies \cite{perkins2013two,perkins2017potential} have explored the potential of applying an external magnetic field along the cylindrical ($z$) axis to mitigate RTI in ICF targets. The RTI along the cylindrical direction in these studies is analogous to the vertical field case in a Cartesian coordinate system, whereas the RTI along the radial ($r$) direction is similar to the parallel field case. The full three-dimensional turbulent mRTI with vertical magnetic fields has been investigated in the Boussinesq incompressible limit \cite{Briard2022PRE,Briard2024JFM,SinghPal2025JFM}. These studies have observed an increasing trend in the mixing width growth with increasing magnetic field strength, which contrasts with the decreasing trend reported for large parallel magnetic fields by Carlyle and Hillier \cite{carlyle2017non}. This difference suggests that distinct dominant mechanisms are at play in the two configurations.

To reconcile the aforementioned scattered findings on mRTI, we conduct a systematic parametric study of the nonlinear regime using both two- and three-dimensional numerical simulations. The effects of external magnetic fields oriented either parallel or perpendicular to the initial interface are examined by varying the magnetic field strength over a broad range. To access the deep nonlinear stage, simulations are performed with aspect ratios up to 8. Our results show that, in both 2D and 3D, weak parallel magnetic fields generally promote the growth of bubbles and spikes, whereas strong parallel fields suppress their development. For vertical magnetic fields, suppression is observed at early times in both 2D and 3D, followed by enhanced growth at later stages. The 2D results agree with those of Jun et al. \cite{jun1995numerical}, while the 3D parallel-field results are consistent with Carlyle and Hillier \cite{carlyle2017non}. The vertical-field cases are in agreement with the findings of Refs.~\cite{Briard2024JFM, SinghPal2025JFM}.

The paper is organized as follows. Section \ref{sec:sim} describes numerical methods. Section III presents the external magnetic effects in 2D and 3D. At last, we conclude and discuss some future work in section IV.

\begin{figure*}
\centering
\begin{subfigure}{0.98\textwidth}
\includegraphics[width=6.2 in]{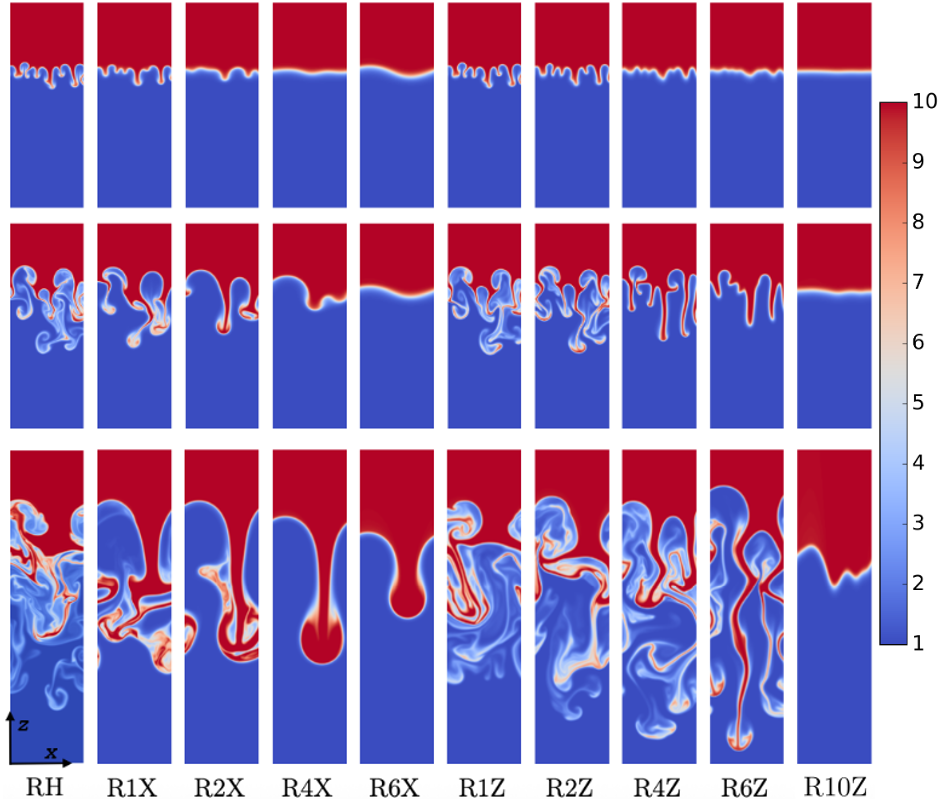}
\end{subfigure}
\renewcommand{\figurename}{Figure}\\
\caption{Visualization of density at $Agt^2/L = 3.9$ (top panels), 10 (middle panels), and 30 (bottom panels) of 2D simulations. Note we have cropped the plots vertically for a better presentation.
}
\label{2D_density_plots}
\end{figure*}

\section{Numerical simulations}\label{sec:sim}

The numerical simulations in this work are performed using the in-house pseudospectral code DiNuSUR, which has previously been applied to hydrodynamic RTI studies \cite{zhao2018inviscid,zhang2018self,Zhao23PoF,ZhaoAluie23PRF}, compressible turbulence \cite{Lees:2019hn}, and incompressible magnetohydrodynamic (MHD) turbulence \cite{bian2019decoupled}. The code has been extended to solve the compressible MHD equations with explicit viscosity and resistivity in Cartesian coordinates \cite{Goedbloed2004}. The governing equations include the continuity equation (\ref{eq:continuity}), the species transport equation (\ref{eq:species}), the momentum equation (\ref{eq:momentum}), the induction equation (\ref{eq:induction}), and the total energy equation (\ref{eq:total}):
\begin{eqnarray}  \label{eq:navier-stokes}
&\displaystyle \frac{\partial \rho}{\partial t}+\grad\bdot(\rho \mathbf{u}) = 0 \label{eq:continuity},\\
&\displaystyle \frac{\partial \rho Y}{\partial t} + \nabla\cdot(\rho Y\mathbf{u}) = \nabla\cdot(\rho D\nabla Y) \label{eq:species},\\
&\displaystyle \frac{\partial \rho\mathbf{u} }{\partial t}+\grad \bdot(\rho \mathbf{uu} - \mathbf{BB}) + \grad p_{tot} =\rho \mathbf{g} + \grad\bdot \mathbf{\sigma} \label{eq:momentum},\\
&\displaystyle \frac{\partial \mathbf{B} }{\partial t} - \grad \btimes ( \mathbf{u}\btimes\mathbf{B}) =  - \grad \btimes (\eta \mathbf{J})  \label{eq:induction},\\
&\displaystyle \frac{\partial (\rho E_{tot})}{\partial t}+\grad \bdot [(\rho E_{tot} + p_{tot})\mathbf{u} - (\mathbf{u}\bdot\mathbf{B})\mathbf{B}]\nonumber\\ & =  \rho \mathbf{u}\bdot \mathbf{g} -\grad\bdot \mathbf{q}  + \grad \bdot [\mathbf{B}\btimes(\eta \mathbf{J})+\mathbf{u}\bdot \mathbf{\sigma}]  \label{eq:total}, \\
&\displaystyle  \grad \bdot \mathbf{B} = 0 \label{eq:divergence-B},
\end{eqnarray}
where $\rho$ denotes the fluid density, $\mathbf{u}$ the fluid velocity, $Y$ the mass fraction of the heavy fluid, and $D$ the mass diffusivity. The total pressure is defined as $p_{tot}=p+|\mathbf{B}|^2/2$, where $p$ is the hydrodynamic pressure. $\mathbf{g}$ is the gravitational acceleration directed along the negative vertical $z$‑axis, the current density is $\mathbf{J} = \nabla \times \mathbf{B}$, and $E_{\mathrm{tot}}$ is the specific total energy per unit mass, defined by 
\begin{eqnarray}
E_{tot}  = e_{int} + |\mathbf{u}|^2/2 + |\mathbf{B}|^2/(2\rho),
\end{eqnarray}
with specific internal energy $e_{int}=c_v T$, the viscous stress $\mathbf{\sigma}_{ij}$ is defined as
\begin{equation}
\sigma_{ij} = 2\mu (S_{ij}-\frac{1}{3}S_{kk} \delta_{ij}),
\end{equation}
where $S_{ij} = (\partial_ju_i+\partial_iu_j) /2$ is the strain‑rate tensor, and $\mu$ is the dynamic viscosity. The heat flux is given by $\mathbf{q} = -\kappa \nabla T$, where $T$ denotes temperature. The system is closed using the ideal‑gas equation of state, $p=\rho R T$ with the specific gas constant $R = 0.8$ in the present simulations. Plasma transport properties — including dynamic viscosity $\mu$, magnetic resistivity $\eta$, thermal conductivity $\kappa$, and specific heats at constant volume $c_v$ and constant pressure $c_p$ — are treated as temporally and spatially uniform constants. The specific‑heat ratio $\gamma = c_p / c_v = 5/3$ is adopted throughout this study. Note that the species transport equation (\ref{eq:species}) describes the evolution of the mass fraction of the heavy fluid, effectively functioning as a passive scalar equation. This formulation is employed exclusively for post‑processing the mixing‑zone growth and does not alter the physical dynamics of the mRTI.

To enforce the non-trivial solenoidal constraint $\nabla\cdot\mathbf{B}=0$ in Eq.~\eqref{eq:divergence-B}, we employ a diffusive cleaning method \cite{marder1987method,van2007hybrid,goedbloed2010advanced,zhang2016comparative}. Specifically, a source term $C_d\,\Delta x^2\,\nabla(\nabla\cdot\mathbf{B})$ is added to the induction equation \eqref{eq:induction}, and an additional term $C_d\,\Delta x^2\,\mathbf{B}\cdot \nabla(\nabla\cdot\mathbf{B})$ is included in the total energy equation \eqref{eq:total}, where $C_d$ is a constant of order unity and $\Delta x$ denotes the grid spacing \cite{goedbloed2010advanced}. The governing equations \eqref{eq:continuity}–\eqref{eq:total} are discretized using a sixth‑order compact finite‑difference scheme \cite{lele1992compact} in the vertical $z$ direction, and a pseudo‑spectral method \cite{canuto2012spectral} with the 2/3-dealiasing rule \cite{PattersonOrszag71} in the horizontal ($x$ and $y$) directions. Time integration is performed with a classical fourth‑order Runge–Kutta scheme. A 2D single‑mode verification shows good agreement with the linear stability analysis of mRTI reported in Refs.~\cite{chandrasekhar1961hydrodynamic,priest2014magnetohydrodynamics} (see Fig.~\ref{fig:lsa}).

We perform both two‑ and three‑dimensional numerical simulations, with an initially uniform magnetic field oriented either along the $x$- or $z$-direction in the mRTI configuration. The initial densities are set to $\rho_l = 1$ for the light fluid and $\rho_h = 10$ for the heavy fluid, corresponding to an Atwood number $A=(\rho_h-\rho_l)/(\rho_h+\rho_l)=0.82$. The initial pressure profile satisfies hydrostatic equilibrium, with the sound speed of the light fluid at the interface normalized to unity. The gravitational acceleration is fixed at $g = 0.1$. The dynamic viscosity $\mu$ is chosen based on numerical convergence tests (see Section~\ref{appendix:convergence_study} and Table~\ref{Tbl:Simulations} in the Appendix). The magnetic Prandtl number is set to $\mathrm{Pr}_m = \nu / \eta \in \{ 1,~0.2,~5 \}$, where the initial kinematic viscosity is defined as $\nu = 2 \mu / (\rho_l + \rho_h)$. The thermal Prandtl number is fixed at $\mathrm{Pr} = c_p \mu / \kappa = 1$.

The simulations are performed with aspect ratios of 4 or 8, ensuring that the vertical extent of the domain is sufficiently large to minimize boundary effects on mixing-layer growth and to accommodate late-time, deep-penetration evolution. In most cases, the computational domain is $L_x \times L_z = 0.1 \times 0.4$ in 2D, and $L_x \times L_y \times L_z = 0.1 \times 0.1 \times 0.4$ in 3D. The corresponding grid resolutions are $N_x \times N_z = 512 \times 2048$ (2D) and $N_x \times N_y \times N_z = 256 \times 256 \times 1024$ (3D). Extended 2D runs, from 2DR1Z to 2DR10Z, employ a taller domain of $0.1 \times 0.8$ with a resolution of $512 \times 4096$. A complete list of simulation parameters is provided in Table~\ref{Tbl:Simulations} in the Appendix. For convenience, simulations are labeled according to their dimensionality, initial magnetic-field strength, and magnetic-field orientation. For example, 2DR6X refers to a two-dimensional run with an initial magnetic field oriented along the $x$-direction with strength of $0.6\,B_c$, where $B_c$ is the critical parallel-field strength from the linear stability analysis in Eq.~\eqref{eq:critical_B}. The primary analysis focuses on cases with unity magnetic Prandtl number ($\mathrm{Pr}_m = 1$), while additional runs with $\mathrm{Pr}_m = 0.2$ and $5$ are included for comparison.

Initially, the upper half of the domain is filled with a uniform heavy fluid of density $\rho_h$, and the lower half with a light fluid of density $\rho_l$. Thus the mass fraction $Y = 1$ in the upper half. The initial pressure field satisfies the hydrostatic condition $ \frac{dP}{dz} = -\rho g $, while the temperature field is obtained from the ideal gas equation of state. The initial perturbation is imposed in the vertical velocity field. In two dimensions, it is given by $v_p(x,z)=\sum_{m} v_{pk}\cos(mk_lx+\phi_{k0})$ and in three dimensions by $v_p(x, y,z)=\sum_{m,n} v_{pk}[\cos(mk_lx+nk_ly+\phi_{k0})]$ where $k_l = 2\pi / L_x = 2\pi / L_y$, and $m$ and $n$ are the mode numbers in the $x$ and $y$ directions, respectively. The phase $\phi_{k0}$ is a random number. The amplitude decreases in $z$ as $v_{pk}=v_{pk0}\exp(-k^2|z-z_0|^2/\pi^2)$ where $k = k_l \sqrt{m^{2} + n^{2}}$ in 3D (and $k = k_l m$ in 2D), $z_0$ is the location of the initial interface, and $v_{pk0}$ is the base amplitude that scales as $k^{-1}$. 
A comparison between Gaussian and exponential decays in $z$ is provided in Appendix~\ref{appendix:IC_compare}.

The detailed influence of initial conditions is beyond the scope of the present study, for discussions on their effects in hydrodynamic RTI, see, e.g., Refs.~\cite{dimonte2004comparative,ramaprabhu2005numerical}. In the initial condition adopted here, the maximum amplitude of the perturbation velocity is around 0.002. The Mach number is defined as $Ma = v_{\mathrm{mag}}/C_s$, where $v_{\mathrm{mag}}$ is the velocity magnitude and $C_s = \sqrt{\gamma P / \rho}$ is the local sound speed. The maximum Mach number increases as the mixing width grows, and, at the same non‑dimensional time, the Mach numbers in the hRTI case are generally higher than those in the mRTI cases..  The maximum Mach numbers reach 0.38 in case 2DRH (at $A g t^{2} / L = 30$) and 0.35 in case 3DRH (at $A g t^{2} / L = 14.7$), respectively. The corresponding root‑mean‑square Mach numbers are 0.097 (2DRH) and 0.054 (3DRH) at these times.  In the initial perturbation spectrum, wavenumbers span $m, n = 1$ to $64$, ensuring that the instability can develop in 2D cases with strong horizontal magnetic fields. Periodic boundary conditions are applied in the horizontal directions ($x, y$). In the vertical direction ($z$), non‑slip wall boundaries are used for velocity, while temperature, species concentration, and magnetic field satisfy zero‑gradient boundary conditions.

In the following analysis, the bubble height $h_b$ is defined as the highest vertical location at which the horizontally averaged mass fraction $\langle Y \rangle_z$ reaches 0.99. Similarly, the spike height $h_s$ is taken as the lowest vertical location where $\langle Y \rangle_z$ reaches 0.01. The mixing layer width is then given by $h = h_b - h_s$. Tests with alternative thresholds, such as 0.05 and 0.95, yield quantitatively similar results. To characterize the degree of internal mixing, we also employ an integral definition of the mixing‑zone width,  
$h_\mathrm{mix} = 3 \int \langle Y \rangle_{xy} \left[ 1 - \langle Y \rangle_{xy} \right] dz$ \cite{Gauthier17}, which emphasizes the mixedness within the mixing region.  

\section{Results and Analysis}

\begin{figure*}
\centering
\begin{subfigure}{0.98\textwidth}
\includegraphics[width=6.2 in]{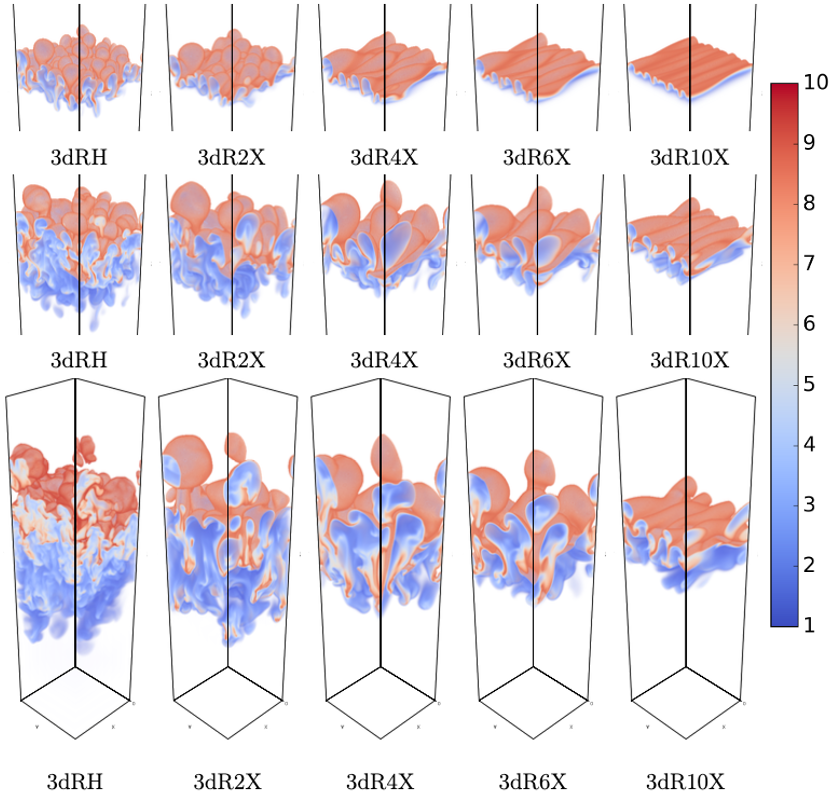}
\end{subfigure}
\renewcommand{\figurename}{Figure}\\
\caption{Visualization of density at $Agt^2/L = 3.9$ (top panels), 8.3 (middle panels) and 14.7 (bottom panels) in 3D hRTI and mRTI with parallel B-fields. Note the plots are cropped for a better presentation.
}
\label{3dVis_X}
\end{figure*}

\begin{figure*}
\centering
\begin{subfigure}{0.98\textwidth}
\includegraphics[width=6.2 in]{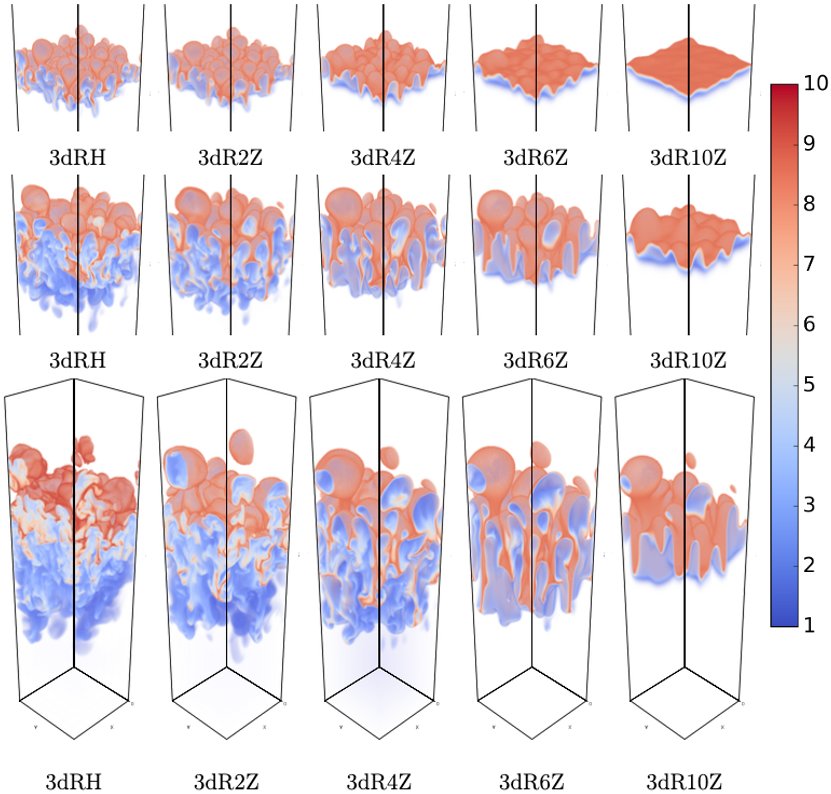}
\end{subfigure}
\renewcommand{\figurename}{Figure}\\
\caption{Visualization of density at $Agt^2/L = 3.9$ (top panels), 8.3 (middle panels) and 14.7 (bottom panels) in 3D hRTI and mRTI with vertical B-fields. Note the plots are cropped for a better presentation.
}
\label{3dVis_Z}
\end{figure*}

We now present the results of the effects of external magnetic fields on mRTI. Initially, we impose a uniform magnetic field with varying strengths $B_0$ in either the parallel ($x$ direction) or vertical ($z$ direction) directions, with the $z$ direction aligned with gravity. We compare these mRTI results to those of hRTI simulations. In both 2D and 3D cases, weak parallel magnetic fields ($B_0 \leq 0.2 B_c$) enhance the development of both bubbles and spikes, whereas stronger magnetic fields reverse this trend. In contrast, vertical magnetic fields initially delay the growth of mRTI, with the delay being nearly monotonic with respect to the field strength. At later times, except possibly for the very large magnetic fields considered here ($B_0=B_c$), larger initial vertical magnetic fields enhance the growth of both bubbles and spikes monotonically. We also observe an asymmetric effect between bubbles and spikes growth due to the large Atwood number (0.82) adopted here. The role of magnetic fields in mRTI nonlinear evolution is analyzed and discussed in the following sections.

\subsection{mRTI structure}
Figure~\ref{2D_density_plots} (2DR1X-2DR6X) and Figure~\ref{3dVis_X} illustrate the density visualizations of mRTI under the influence of parallel magnetic fields in 2D and 3D, respectively, at various time instants $A g t^2/L$, where $ L = L_x$. Both 2D and 3D results indicate that parallel magnetic fields suppress small-scale structures and increase bubble size. Given that the terminal velocity of bubbles in turbulent RTI is proportional to the square root of their size \cite{Oron01,Goncharov02}, the increased size of horizontal mRTI bubbles may enhance the growth rate of the mixing zone. In 2D, for strong magnetic fields (2DR4X and 2DR6X), only modes proportional to $L$ survive, and the development resembles that of single-mode RTI at late times. In 3D, strong magnetic fields (3DR4X-3DR10X) generate anisotropic structures in the horizontal ($x-y$) plane at $A g t^2/L = 3.9 $ and 8.3. The undular modes, depicted in the $x-z$ plane, are significantly suppressed by the external magnetic field in the $x$ direction, while the interchange modes, shown in the $y-z$ plane, resemble hRTI but with a time delay. By $A g t^2/L= 14.7 $, the anisotropic structures in the $x-y$ plane become less pronounced, except in the case of 3DR10X.

Figure~\ref{2D_density_plots} (2DR1Z–2DR6Z) and Figure~\ref{3dVis_Z} show density visualizations of mRTI in the presence of vertical magnetic fields for the 2‑D and 3‑D cases, respectively. The results indicate that vertical fields suppress small‑scale structures less effectively than parallel fields. In the vertical‑field cases, spikes become elongated and develop into collimated, thread‑like structures, while horizontal structures are largely inhibited. The resulting flow anisotropy differs qualitatively from that produced by parallel magnetic fields.

\subsection{Mixing width and energy evolution}

We now examine the time evolution of the bubble and spike heights, defined by the 0.01-0.99 threshold of $\langle Y\rangle_{xy}$, as well as the integral mixing zone width, defined by $h_\mathrm{mix}=3\int \langle Y\rangle_{xy} (1-\langle Y\rangle_{xy}) dz$ \cite{Gauthier17}, where $\langle\cdot\rangle_{xy}$ denotes horizontal averaging. The bubble and spike heights delineate the upper and lower envelop of the mixing zone, while the mixing-zone width $h_{\text{mix}}$ is an integrated measure that highlights the extent of mixing within the mixing zone. Overall, weak horizontal magnetic fields promote both bubble and spike growth but stronger ones suppress them, whereas vertical magnetic fields, after an initial transient, tend to enhance the growth of the mixing zone.

\begin{figure*}
\centering
\begin{subfigure}{0.99\textwidth}
\includegraphics[width=6 in]{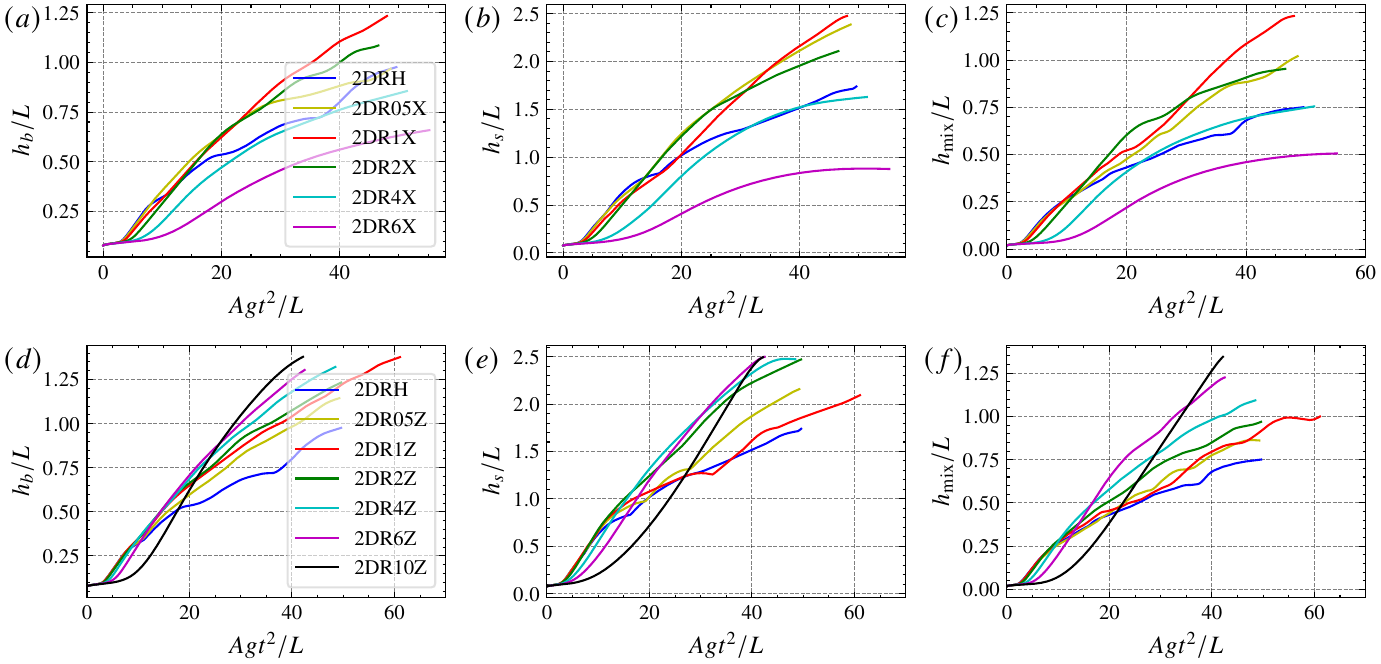}
\end{subfigure}
\renewcommand{\figurename}{Figure}\\
\caption{Time evolution of the bubble height $(a,d)$, spike height $(b,e)$, and mixing zone width $(c,f)$ for 2D simulations with $\mathrm{Pr}_m=1$. Panels (a–c) illustrate the effects of horizontal magnetic fields, whereas panels (d–f) show the effects of vertical magnetic fields.
}
\label{fig:2Dmixingheight}
\end{figure*}

\subsubsection{2D mixing-zone width}

Figure~\ref{fig:2Dmixingheight}(a-c) shows the time evolution of the 2D bubble, spike, and mixing zone heights, respectively, under the influence of parallel magnetic fields. In panels (a) and (b), weak parallel fields ($B_{0} \leq 0.2\,B_{c}$) enhance both bubble and spike growth relative to the hydrodynamic RTI case (2DRH). As the field strength increases (cases 2DR4X–2DR6X), the growth of both structures is progressively stabilized compared with 2DRH. For 2DR6X, the parallel field is sufficiently strong that mRTI development is almost completely suppressed at late times $(A g t^{2} / L > 50$). The mixing zone width, $h_{\mathrm{mix}}$ in panel (c), exhibits trends similar to those of the bubble and spike. Comparing different horizontal mRTI cases, the relative magnitudes of bubble, spike, and mixing-zone growth all increase with weak fields and decrease monotonically with stronger $B_{0}$.

Figure~\ref{fig:2Dmixingheight}(d–f) shows the time evolution of the bubble, spike, and mixing‑zone heights in two‑dimensional mRTI with vertical magnetic fields. The evolution proceeds through three characteristic phases. In the initial delaying stage ($A g t^{2} / L < 10$), as predicted by the linear stability analysis in Eq.~\eqref{eq:vB_growth}, vertical fields impose a monotonic stabilizing effect on both bubbles and spikes, suppressing their early growth, with stronger fields extending this phase. This is followed by an acceleration stage in which both bubbles and spikes grow more rapidly, with growth rates increasing with magnetic‑field strength. This trend is consistent with previous findings \cite{Grea2023APJ} that terminal bubble and spike velocities rise monotonically with the strength of an external vertical field. Density visualizations in Fig.~\ref{2D_density_plots} show that stronger fields produce larger bubbles and more vertically elongated spikes, with kinetic energy predominantly aligned in the vertical direction, enhancing their rise and fall speeds. For sufficiently strong fields, the heights of both bubbles and spikes ultimately surpass those in the hydrodynamic case and continue to increase monotonically with field strength. At later times, interactions between multiple bubbles and spikes generate significant horizontal motion, marking the onset of the self‑similar regime in which vertical growth slows and follows an approximately quadratic‑in‑time scaling (which appears linear when plotted against $A g t^{2} / L$). For strong vertical fields, this transition is significantly delayed or may not occur within the simulated time span. This allows vertical growth to remain higher than in weak‑field and hydrodynamic cases until the end of the simulation.

The 2D results presented here are consistent with those of Jun \etal~\cite{jun1995numerical}, though obtained at a larger aspect ratio and higher Reynolds numbers than their results. These findings also carry implications for National Ignition Facility (NIF) target design. Perkins \etal~\cite{perkins2017potential} proposed applying an axial magnetic field to improve NIF performance. However, our results indicate that in 2D, although strong parallel fields have a stabilizing effect, strong vertical fields can have destabilizing effects. Consequently, both the strength and orientation of the applied magnetic field must be carefully optimized to control instability growth.

\subsubsection{3D mixing width}

\begin{figure*}
\centering
\begin{subfigure}{0.99\textwidth}
\includegraphics[width=6 in]{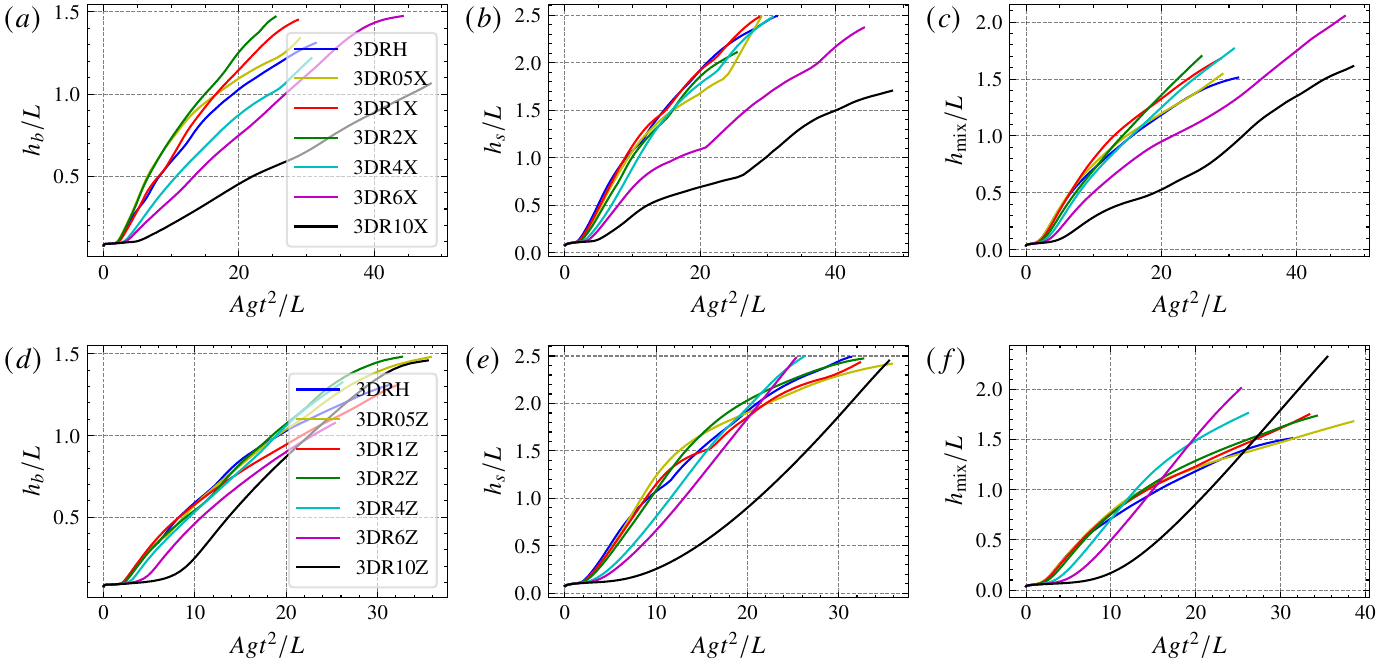}
\end{subfigure}
\renewcommand{\figurename}{Figure}\\
\caption{Time evolution of the bubble height $(a,d)$, spike height $(b,e)$, and mixing zone width $(c,f)$ for 3D simulations with $\mathrm{Pr}_m=1$. Panels (a–c) illustrate the effects of horizontal magnetic fields, whereas panels (d–f) show the effects of vertical magnetic fields.
}
\label{fig:3Dmixingheight}
\end{figure*}

Figures~\ref{fig:3Dmixingheight}(a–c) present the time evolution of the bubble height, spike height, and mixing width in 3D under parallel magnetic fields. In Fig.~\ref{fig:3Dmixingheight}(a), as in the 2D case, weak parallel fields enhance bubble growth, whereas strong parallel fields suppress it. Figure~\ref{fig:3Dmixingheight}(b) shows that spike development remains largely unchanged for $B_{0} \leq 0.4\,B_{c}$, but stronger fields effectively suppress spike growth. The mixing width trends in Fig.~\ref{fig:3Dmixingheight}(c) further indicate that, relative to the hRTI case, horizontal fields with $B_{0} < 0.2\,B_{c}$ enhance mixing zone growth, while $B_{0} \geq 0.2\,B_{c}$ leads to progressive suppression with increasing $B_{0}$.

Figures~\ref{fig:3Dmixingheight}(d–f) show the corresponding results for vertical magnetic fields. Compared with the 2D results in Figs.~\ref{fig:2Dmixingheight}(d,e), the 3D growth rates are faster because the simulation includes the longest-wavelength mode ($k = 1$), whose terminal bubble and spike velocities are greater in 3D than in 2D \cite{Goncharov02}. As in 2D, the bubble evolution in Fig.~\ref{fig:3Dmixingheight}(d) proceeds through an initial delaying stage followed by an acceleration stage. However, the acceleration is weaker than in 2D, so that at late times only weak vertical-field cases surpass the hRTI case, while strong-field cases may require a greater vertical extent to overtake it. 
Spike growth in Fig.~\ref{fig:3Dmixingheight}(e) shows analogous delaying, acceleration, and self‑similar growth stages. At late times, stronger vertical fields generally yield greater spike heights, with the exception of the $B_{0} = B_{c}$ case, which remains consistently lower. The total mixing width in Fig.~\ref{fig:3Dmixingheight}(f) exhibits a trend similar to that of the spikes, but the self‑similar stage is observed only for field strengths $B_0 \leq 0.4 B_c$. For stronger fields, self‑similarity is absent, which can be attributed to the suppression of horizontal interactions.

We now compare our 3D mRTI results with previous studies on parallel magnetic fields \cite{stone2007magnetic,stone2007nonlinear,carlyle2017non,Kalluri2025JFM,BriardGrea2025} and on vertical magnetic fields \cite{Briard2022PRE,Briard2024JFM,SinghPal2025JFM}. Carlyle and Hillier \cite{carlyle2017non} performed only mRTI simulations (i.e., without comparison to hRTI) using parallel fields no weaker than $0.6B_{c}$, and found that increasing field strength generally reduced the mRTI growth rate, which is consistent with our observations. However, the domain aspect ratio used in Ref.~\cite{carlyle2017non} (1:0.25:4, with the magnetic field aligned along the first dimension) constrained the growth of horizontal interchange modes. In contrast, our visualizations in Fig.~\ref{3dVis_X} show no evidence of anisotropy in the $x$–$y$ plane at late times for cases 3DR1X–3DR6X, a feature absent in \cite{carlyle2017non}. We emphasize that a sufficiently large domain in both the $x$ and $y$ directions is also essential for achieving self-similar RTI growth. In addition, Kalluri and Hillier \cite{Kalluri2025JFM} performed incompressible, variable-density mRTI simulations at relatively weak parallel fields ($\leq 0.25\,B_c$) in 3D and reported a monotonically increasing mixing-layer growth. This trend is consistent with our results shown in Figs.~\ref{fig:3Dmixingheight}(c). Similarly, Briard et al. \citep{BriardGrea2025}, in a Boussinesq framework, demonstrated enhanced mixing‑layer growth for parallel magnetic fields up to $0.42 B_c$ relative to the hRTI case, which is also consistent with our findings.

Refs.~\cite{stone2007magnetic,stone2007nonlinear} reported that parallel magnetic fields enhance bubble development even for a relatively strong field of $0.6\,B_c$, which contradicts our finding that, at $B_0 = 0.6\,B_c$, parallel mRTI evolves more slowly than hRTI. However, their observation that parallel mRTI with a weaker field ($0.2\,B_c$) grows faster than in the $0.6\,B_c$ case is consistent with our results.  
A possible explanation for the discrepancy is that the simulations in \cite{stone2007magnetic,stone2007nonlinear} employed an ideal-MHD formulation without explicit physical viscosity or magnetic diffusivity, which can make numerical convergence less certain. As demonstrated in Fig.~\ref{appendix:2dConverge} in the Appendix, under-resolved simulations can lead to qualitatively different conclusions. Although \cite{stone2007magnetic,stone2007nonlinear} included a convergence study, it was limited to 2D single-mode cases and covered only the linear to early nonlinear stages. In contrast, we performed convergence tests using 2D multi-mode configurations evolved to the deep nonlinear stage, which offers greater confidence in the robustness of our results.  
We also note that hRTI is generally more sensitive to grid resolution, as magnetic tension in mRTI suppresses small-scale features and thus relaxes resolution requirements. This difference in resolution sensitivity may also contribute to the variation in hRTI growth rates between \cite{stone2007magnetic,stone2007nonlinear} and the present study.

For vertical magnetic fields, Refs.~\cite{Briard2022PRE,Briard2024JFM,SinghPal2025JFM}, using incompressible Boussinesq formulations, reported that the mixing‑zone growth increases with field strength—up to $0.71\,B_c$ in Ref.~\cite{Briard2024JFM} and $0.21\,B_c$ in Ref.~\cite{SinghPal2025JFM}. This trend is consistent with our results, where the evolution proceeds through three distinct stages: an initial delaying stage, followed by an acceleration stage, and finally a self‑similar stage (see, for example, Fig.~4(b) in Ref.~\cite{Briard2024JFM}).  
Notably, in Ref.~\cite{Briard2024JFM}, the larger Reynolds number causes the transition from the acceleration stage to the self‑similar stage to occur earlier than in the current study. This is because, at higher Reynolds numbers, vertically rising bubbles and falling spikes interact more readily, generating horizontal motions that slow down vertical growth and thus shorten the acceleration stage.

\subsubsection{Effect of initial perturbations}

The results presented in Figs.~\ref{fig:2Dmixingheight} and \ref{fig:3Dmixingheight} correspond to initial velocity perturbations with wavenumbers $k \in [1, 64]$, representing a broadband perturbation spectrum. This configuration can introduce lateral-confinement effects at late times.  To examine RT cases with minimal lateral-confinement influence, we performed additional 2D and 3D simulations, similar to those listed in Table~\ref{Tbl:Simulations}, but with initial velocity perturbations restricted to $k \in [32, 64]$. These high wavenumber perturbation configurations lie within the bubble-merger regime \citep{ZHOU20171}.  

\begin{figure*}
\centering
\begin{subfigure}{0.99\textwidth}
\includegraphics[width=6 in]{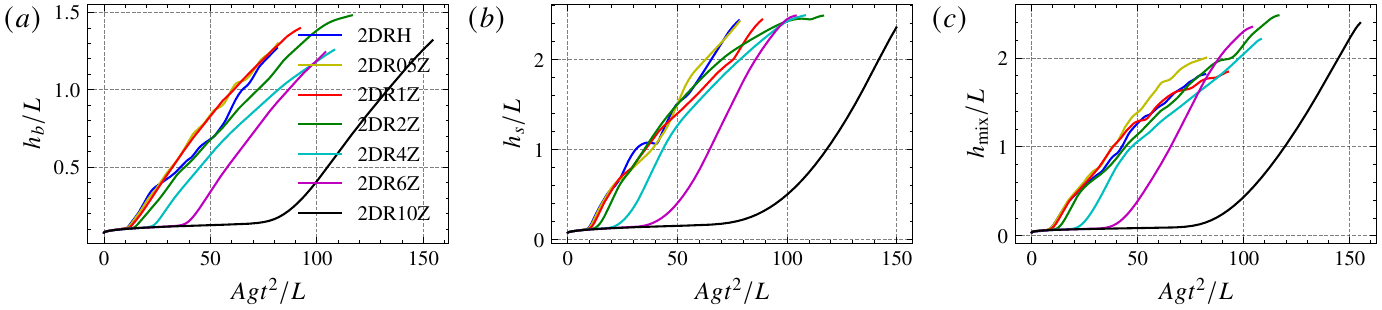}
\end{subfigure}
\renewcommand{\figurename}{Figure}\\
\caption{Time evolution of the bubble height $(a)$, spike height $(b)$, and mixing zone width $(c)$ for 2D hRTI and mRTI simulations with vertical external magnetic field and initial velocity perturbations within the wavenumber range $k\in [32, 64]$. \label{Appfig:2D_width_merger}
}
\end{figure*}

\begin{figure*}
\centering
\begin{subfigure}{0.99\textwidth}
\includegraphics[width=6 in]{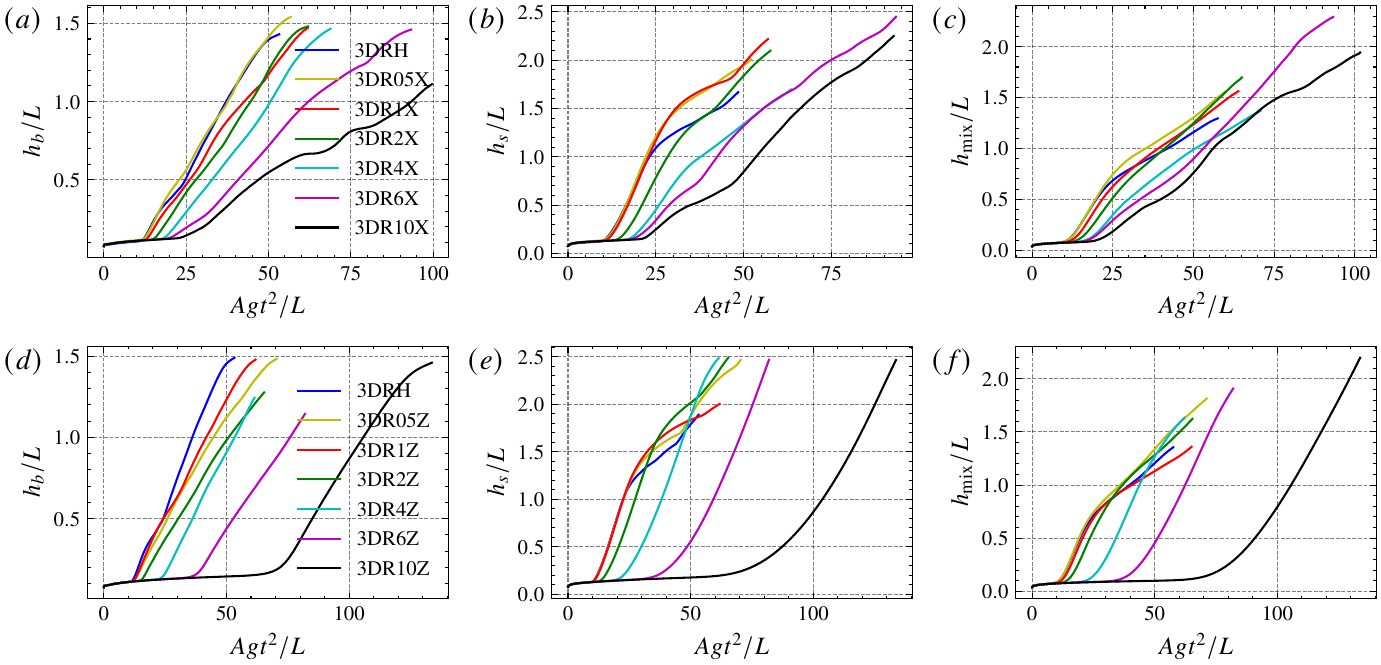}
\end{subfigure}
\renewcommand{\figurename}{Figure}\\
\caption{Time evolution of the bubble height $(a,d)$, spike height $(b,e)$, and mixing zone width $(c,f)$ for 3D simulations with $\mathrm{Pr}_m=1$ and initial velocity perturbations within the wavenumber range $k\in [32, 64]$. Panels (a–c) illustrate the effects of horizontal magnetic fields, whereas panels (d–f) show the effects of vertical magnetic fields. \label{Appfig:3D_width_merger}
}
\end{figure*}

For the 3D simulations, we considered both horizontal and vertical magnetic fields. In the 2D simulations, we examined only vertical fields, as the horizontal-field cases are fully suppressed under such small-scale initial perturbations. Figure~\ref{Appfig:2D_width_merger} presents the bubble height, spike height, and mixing-zone width for the 2D hRTI and vertical mRTI cases, while Fig.~\ref{Appfig:3D_width_merger} shows corresponding results for the 3D cases with both horizontal and vertical external magnetic fields.  

From these figures, we observe that in the bubble-merger regime (high-wavenumber initial perturbations), the simulations require a significantly longer time to reach the same bubble, spike, and mixing-zone heights compared to simulations incorporating the $k=1$ mode shown in figures~\ref{fig:2Dmixingheight} and \ref{fig:3Dmixingheight}. This is because the terminal velocity increases with wavelength in both hRTI and mRTI cases. Nevertheless, the qualitative trends remain consistent: weak horizontal fields enhance mixing-zone growth, whereas strong horizontal fields suppress it; vertical fields delay the onset of mixing-zone growth but may later enhance growth once the system enters the self-similar regime.  

Thus, lateral confinement effects present in broadband perturbation cases do not alter the qualitative trend in the relative growth rates of the mRTI mixing zone across different external-field strengths. Lateral confinement may change the absolute speeds of bubble/spike growth by restricting merging and mode availability, but the relative differences between external field strengths are governed by the physics of mRTI growth, which act on the available modes the same way regardless of confinement.

\subsubsection{Evolution of kinetic and magnetic energies}

\begin{figure*}
\centering
\begin{subfigure}{0.99\textwidth}
\includegraphics[width=6 in]{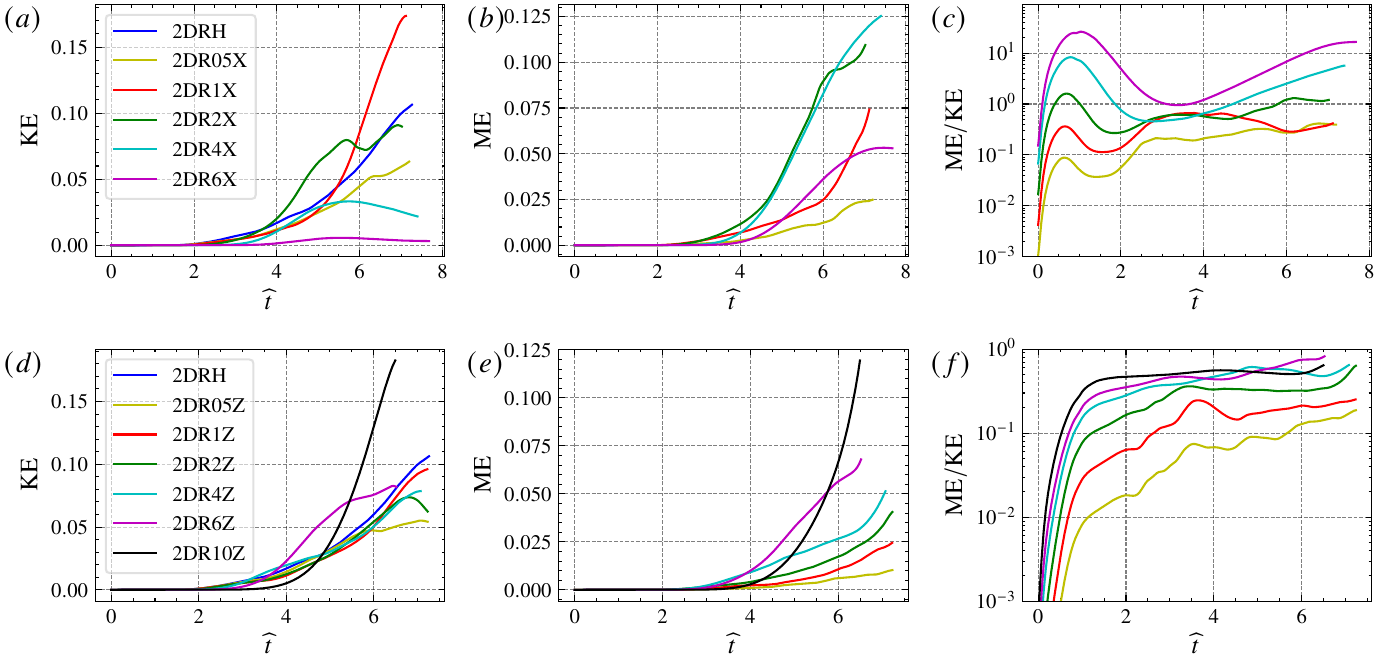}
\end{subfigure}
\renewcommand{\figurename}{Figure}\\
\caption{Time evolution of $(a,d)$ kinetic energy, $(b,e)$ magnetic energy, and $(c,f)$ their ratio from 2‑D simulations at $\mathrm{Pr}_m=1$, versus non-dimensional time $\widehat{t} = \sqrt{A g t^{2} / L}$. Top row: horizontal magnetic field; bottom row: vertical magnetic field. All energies are normalized by $\rho_0 A g L_z/2$, with $\rho_0 = (\rho_h + \rho_l)/2$. Panels (c) and (f) use a logarithmic vertical scale.  
}
\label{fig:2DKEME}
\end{figure*}

\begin{figure*}
\centering
\begin{subfigure}{0.99\textwidth}
\includegraphics[width=6 in]{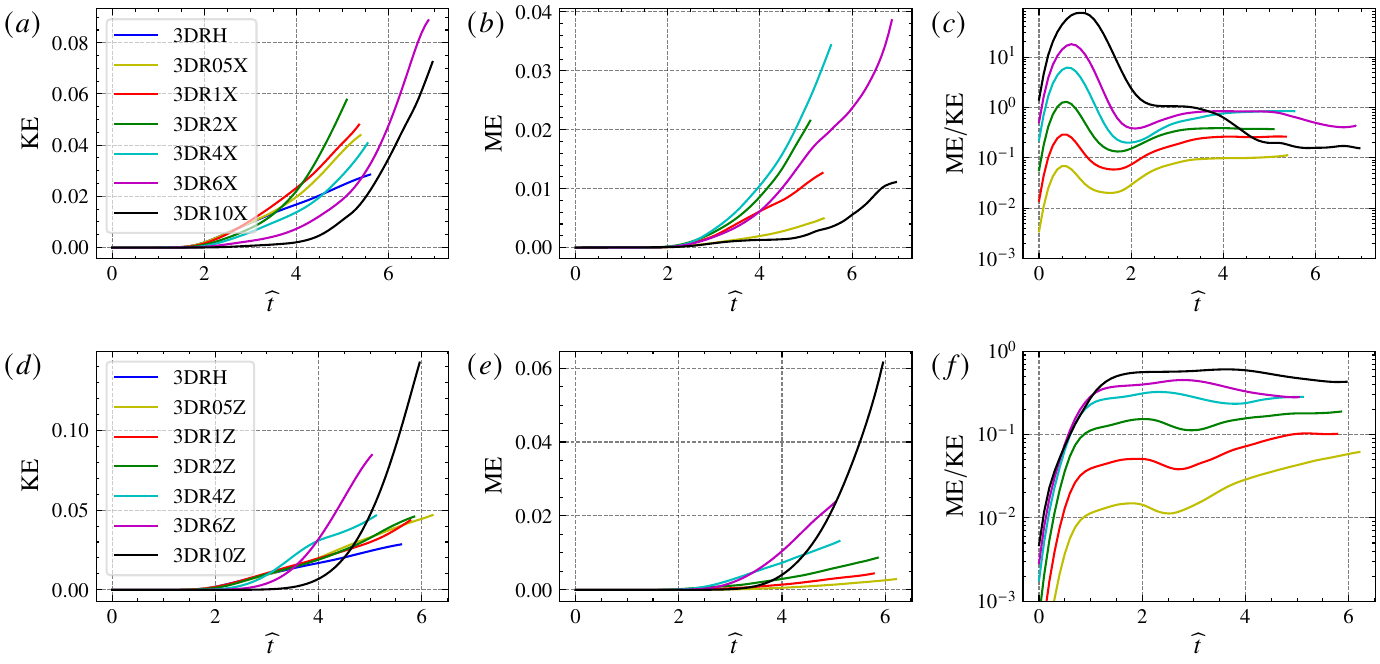}
\end{subfigure}
\renewcommand{\figurename}{Figure}\\
\caption{Time evolution of $(a,d)$ kinetic energy, $(b,e)$ magnetic energy, and $(c,f)$ their ratio from 3‑D simulations at $\mathrm{Pr}_m=1$, versus non-dimensional time $\widehat{t} = \sqrt{A g t^{2} / L}$. Top row: horizontal magnetic field; bottom row: vertical magnetic field. All energies are normalized by $\rho_0 A g L_z/2$, with $\rho_0 = (\rho_h + \rho_l)/2$. Panels (c) and (f) use a logarithmic vertical scale.  
}
\label{fig:3DKEME}
\end{figure*}

The time evolutions of the total kinetic energy ($\mathrm{KE}=\langle \frac{1}{2}\rho |\bu|^2\rangle$), magnetic energy ($\mathrm{ME}=\langle \frac{1}{2}|\bB-\bB_0|^2\rangle$, where $\bB_0$ is the external magnetic field), and their ratio are shown in Fig.~\ref{fig:2DKEME} for the 2D mRTI and in Fig.~\ref{fig:3DKEME} for the 3D cases.  
In the 2D results of Fig.~\ref{fig:2DKEME}, KE for weak horizontal fields fluctuates around the hydrodynamic RTI (hRTI) baseline. When the horizontal field is strong ($B_0 \ge 0.4B_c$), KE is markedly suppressed and can even decrease for $\widehat{t} = \sqrt{A g t^2 / L} > 5$, despite continued growth of the mixing zone. For vertical fields, weak‑field cases again fluctuate around the hRTI level, whereas strong fields ($B_0 \ge 0.6B_c$) produce a net increase in KE.  
The ME evolution in 2D broadly follows that of the mixing‑zone width: for horizontal fields, ME magnitude rises with $B_0$ in the weak‑ to moderate‑field regime but declines when the field becomes strong; for vertical fields, ME growth is enhanced with increasing $B_0$ during the late nonlinear stage.  
In contrast, the 3D results of Fig.~\ref{fig:3DKEME} show that, for both horizontal and vertical fields, the relative trends of KE and ME closely track those of the corresponding mixing‑zone growth.

In mRTI, the initially unstable gravitational potential energy is converted into KE through gravity injection during the growth of the instability, and KE is subsequently transferred to ME via the Lorentz force. The relative magnitude of KE and ME during this process depends strongly on the strength and orientation of the external magnetic field. The time evolution of the ME/KE ratio is shown in Fig.~\ref{fig:2DKEME}(c,f) for the 2‑D simulations and Fig.~\ref{fig:3DKEME}(c,f) for the 3‑D simulations.  

For strong horizontal fields, vertical RTI motions bend magnetic field lines efficiently, enabling rapid ME growth while KE is strongly suppressed by magnetic tension. This leads to $\mathrm{ME}/\mathrm{KE} \gg 1$ in the weakly nonlinear stage ($\widehat{t} < 2$), as seen in Figs.~\ref{fig:2DKEME}(c) and \ref{fig:3DKEME}(c). As the flow evolves and the accumulated tension is partially released, ME growth slows while KE recovers, causing the ratio to decrease toward unity. An exception is observed in the 2‑D strong‑horizontal‑field cases (2DR4X and 2DR6X), where the interchange mode is absent and the flow remains highly regular and dominated by the largest wavelength mode (Fig.~\ref{2D_density_plots}). In these cases, ME continues accumulating with little release, preventing $\mathrm{ME}/\mathrm{KE}$ from approaching unity even at late times.  

In the vertical‑field configuration, the dominant vertical motions are aligned with $B_0$, producing minimal bending of magnetic field lines. Transverse motions, which could otherwise amplify ME, are strongly inhibited. Consequently, $\mathrm{ME}/\mathrm{KE}$ remains below unity for the entire evolution in both 2‑D and 3‑D simulations.

\subsection{Roles of magnetic fields}

We next examine the role of magnetic fields in the evolution of mRTI. Magnetic fields affect the instability primarily in two ways. First, they modify the energy transfer pathways by altering flow anisotropy, which stems from changes in the morphology of the density and velocity fields. Second, they influence interface dynamics through the interplay of physical processes such as the Lorentz force, drag, buoyancy, and vorticity generation, which act differently on bubbles and spikes depending on the magnetic‑field strength.

These effects can be understood by examining the overall force balance on a rising bubble (with similar considerations applying to spikes) under different external magnetic fields. Applying Newton’s second law to a bubble with vertical velocity $u_b$ yields
\begin{align}
\rho_{h} V \partial u_b / \partial t = & \underbrace{V (\boldsymbol{J \times} \bB)_z}_{\text{Lorenz force}} \underbrace{-C_d \rho_{h} u_b^{2} A_h}_{\text{drag}}   \nonumber \\
& + \underbrace{C_b\Delta \rho  V g}_{\text{buoyancy}} + \underbrace{\rho_l V \omega_0^2 R/4}_{\text{vorticity}} 
\label{eq:bub_dynamics},
\end{align}
where $V$ is the volume of the fluid element. The first term on RHS is the $z$-direction Lorenz force (The $x$-component also plays an important in shaping flow anisotropy). The second one is the drag term $F_d = -C_d \rho_{h} u_b^{2} A_h$, where $A_h$ is the area projected onto horizontal plane, and $C_d$ is drag coefficient \cite{takabe1991reduction,alon1995power} of order unity. The third term is the buoyancy term, where $\Delta \rho$ is the density difference and $C_b$ is an efficient coefficient of order unity that accounts for the mixedness between two fluids. The last term is the vorticity term, measuring the centrifugal force exerted by vortices inside the bubble, where $\omega_0 = \frac{\int_V\rho |\bm \omega|^2dV}{2\int_V\rho dV}$ is the spatial average of the vorticity inside the bubble tip, ${\bm \omega} = {\grad \btimes \mathbf{u}}$, and $R$ is the radius of the bubble \cite{betti2006bubble,yan2016three,zhang2018self,xbian2019singlemode,luo2020effects}. Overall, the first two terms on the RHS hinders the motion of bubbles, while the last two terms tend to enhance the bubble motion. We now briefly examine the individual physical quantities relevant to these effects.

\subsubsection{Effect of Lorentz force}

\begin{figure*}
\centering
\begin{subfigure}{0.99\textwidth}
\includegraphics[width=5. in]{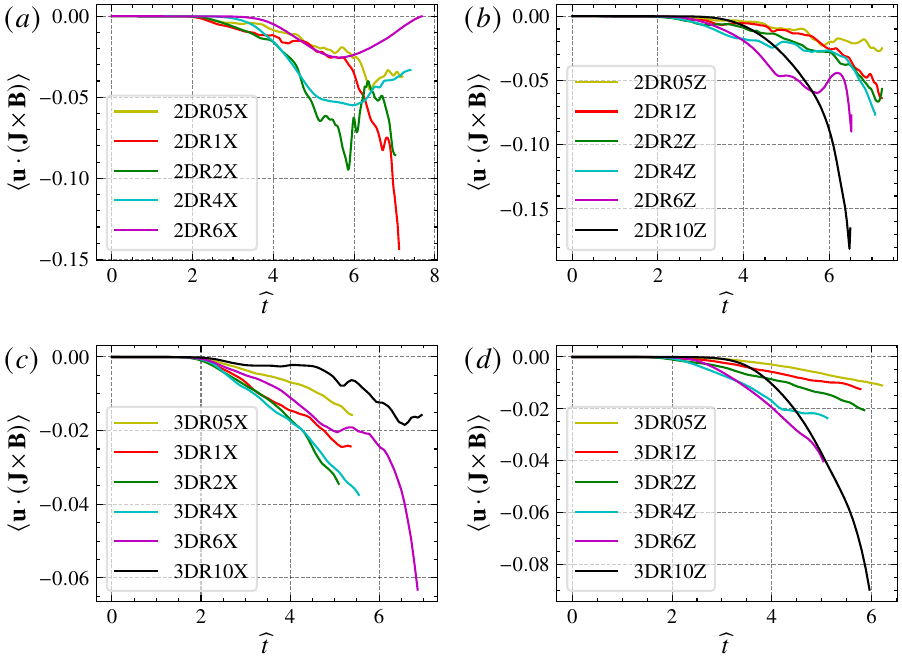}
\end{subfigure}
\renewcommand{\figurename}{Figure}\\
\caption{The mean KE to ME conversion term  $\langle \bu\cdot(\mathbf{J}\times\mathbf{B})\rangle$ for 2D in panels (a,b), and for 3D in panels (c,d). The conversion term is normalized by $\rho_0 (Ag)^{3/2}L^{1/2}$.
}
\label{fig:2D3Dujb}
\end{figure*}

The Lorentz force $(\mathbf{J} \times \mathbf{B})$, with $\mathbf{J} = \nabla \times \mathbf{B}$ the current density, constitutes the most direct dynamical distinction between mRTI and hRTI. It modifies the interface evolution through the combined effects of magnetic tension and magnetic pressure, and alters the anisotropy statistics during the evolution of the instability. In the vertical direction, the Lorentz force acts as a restoring force opposing the motion of bubbles and spikes, while also indirectly influencing the drag, buoyancy, and vorticity terms.  
 
The Lorentz force per unit volume can be decomposed into magnetic pressure and magnetic tension contributions as  
\begin{equation}
\mathbf{J} \btimes \mathbf{B}=(\mathbf{\grad} \btimes \mathbf{B}) \btimes \mathbf{B}=-\grad\left(\frac{|\bB|^{2}}{2 }\right)+(\mathbf{B} \bdot \mathbf{\grad}) \mathbf{B}.
\end{equation}
When dotted with the velocity field, $\mathbf{u} \cdot (\mathbf{J} \times \mathbf{B})$ represents the rate of energy exchange between the kinetic and magnetic energy reservoirs.  

Figure~\ref{fig:2D3Dujb} shows the mean energy conversion term $\langle\mathbf{u} \cdot (\mathbf{J} \times \mathbf{B})\rangle$, which is negative in all cases, indicating a net transfer of KE to ME. For horizontal magnetic fields in both 2D and 3D simulations, the magnitude of this conversion term increases with weak field strength, reaches a maximum around $B_0=0.2B_c$, and then decreases as the field becomes strong. For vertical fields, the conversion magnitude generally increases with field strength, except in the very strong‐field limit, a trend consistent with the corresponding mixing‐zone growth.  
Comparing dimensionality, the conversion term in 2D is approximately twice as large as in 3D. This difference can be attributed to the reduced downscale energy transfer and dissipation in 2D flows due to the absence of vortex stretching, which maintains a higher KE level as well as the conversion term in 2D than in 3D, consistent with the results in Figs.~\ref{fig:2DKEME} and \ref{fig:3DKEME}.

\begin{figure*}
\centering
\begin{subfigure}{0.99\textwidth}
\includegraphics[width=5 in]{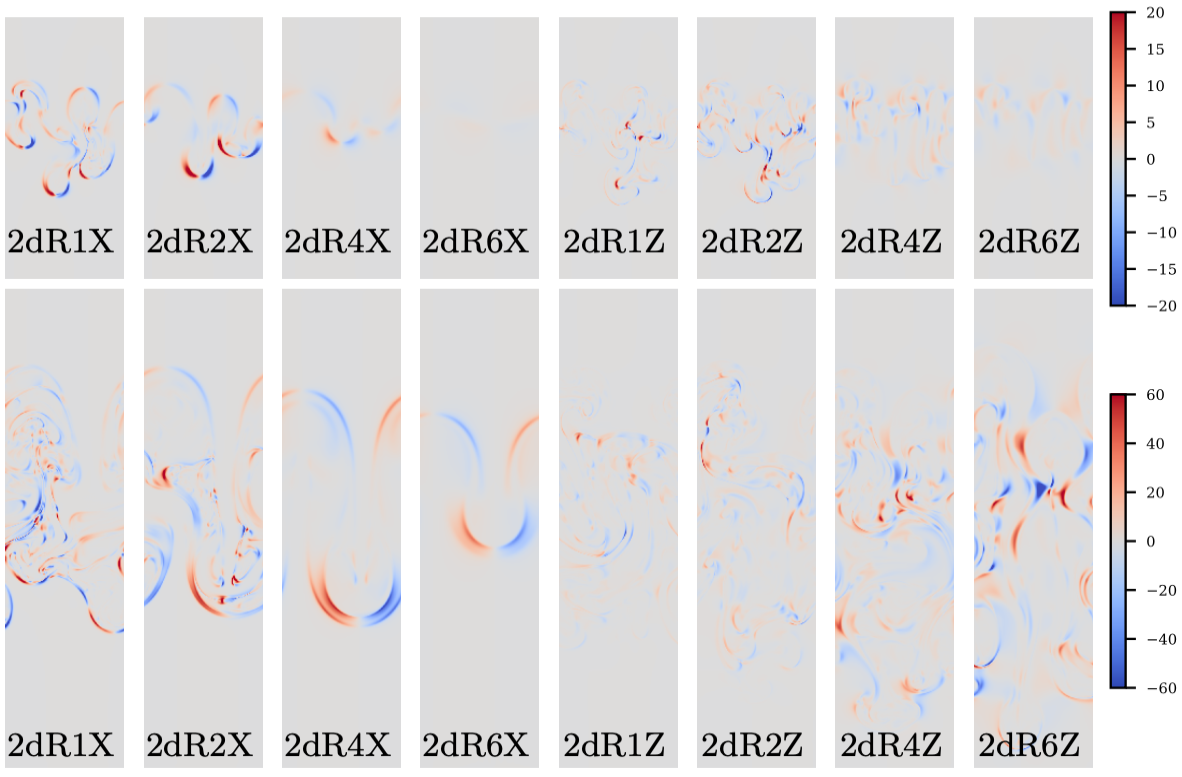}
\end{subfigure}
\renewcommand{\figurename}{Figure}\\
\caption{Visualization of dimensionless magnetic tension $T_x^* = (\bB\bdot\grad \bB)_x / (B_c^2/L)$ at $Agt^2/L = 10$ (top panels) and 30 (bottom panels) of 2D simulations. The magnetic tension is normalized by $B_c^2/L$, where $B_c$ is the critical magnetic field. Note we crop the plots vertically to save space. 
}
\label{fig:2D_Tx_plots}
\end{figure*}

\begin{figure*}
\centering
\begin{subfigure}{0.99\textwidth}
\includegraphics[width=5. in]{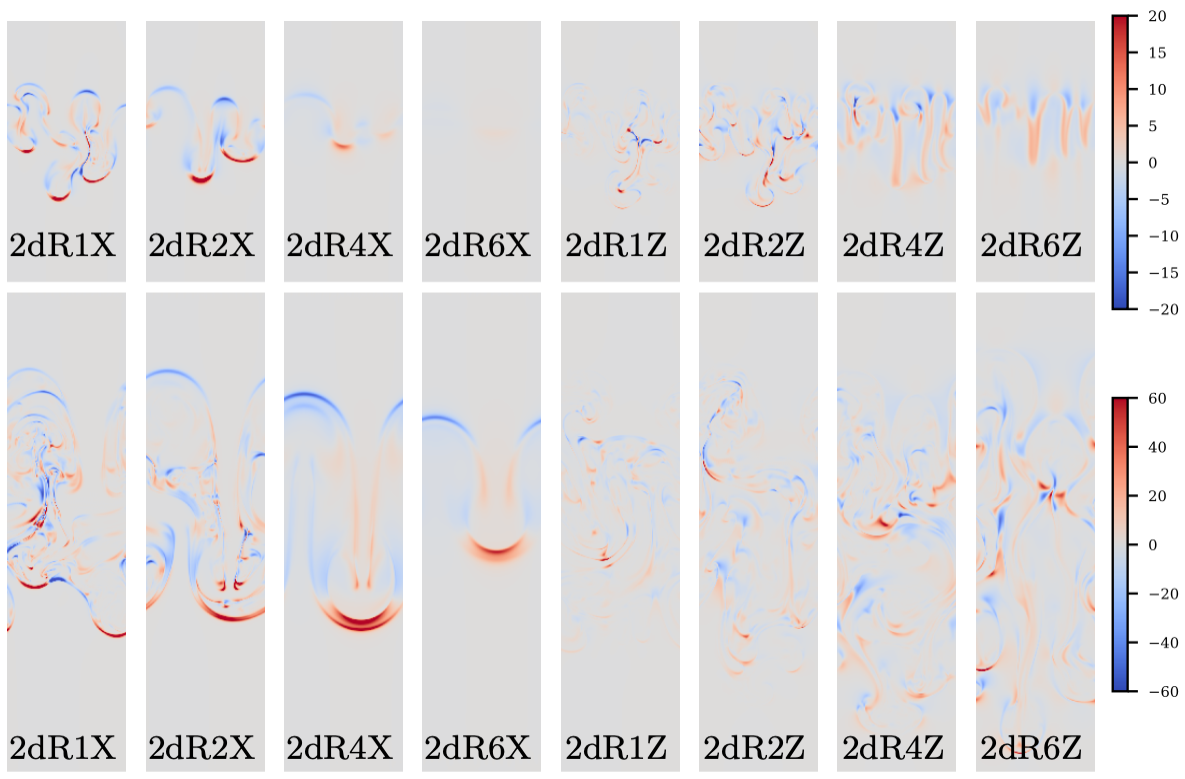}
\end{subfigure}
\renewcommand{\figurename}{Figure}\\
\caption{Visualization of dimensionless magnetic tension $T_z^* = (\bB\bdot\grad \bB)_z / (B_c^2/L)$ at $Agt^2/L = 10$ (top panels) and 30 (bottom panels) of 2D simulations.The magnetic tension is normalized by $B_c^2/L$, where $B_c$ is the critical magnetic field. Note we crop the plots vertically to save space. 
}
\label{fig:2D_Tz_plots}
\end{figure*}

In our mRTI simulations, the flow Mach number is small and the system is nearly incompressible, so this energy exchange is dominated by the magnetic tension term $\bu\cdot(\bB\cdot\nabla\bB)$, while the spatial mean of the magnetic-pressure contribution $\mathbf{u} \cdot \nabla(|\mathbf{B}|^2/2)$ is negligible.  
Notably, local magnetic pressure can still be 5–10 times larger than magnetic tension in magnitude, but its net (integrated) contribution to energy exchange remains small. For this reason, we focus on the role of magnetic tension $\mathbf{T} \equiv (\mathbf{B} \cdot \nabla)\mathbf{B}$. Figures~\ref{fig:2D_Tx_plots} and \ref{fig:2D_Tz_plots} show $T_x$ and $T_z$ for the 2‑D mRTI cases. As expected, $T_x$ suppresses horizontal motions, while $T_z$ opposes vertical displacements, acting as a restoring force. The magnitudes of $T_x$ and $T_z$ vary with magnetic‐field strength, and they exert drastic influence on the mRTI flow anisotropy.

\begin{figure*}
\centering
\begin{subfigure}{0.99\textwidth}
\includegraphics[width=5. in]{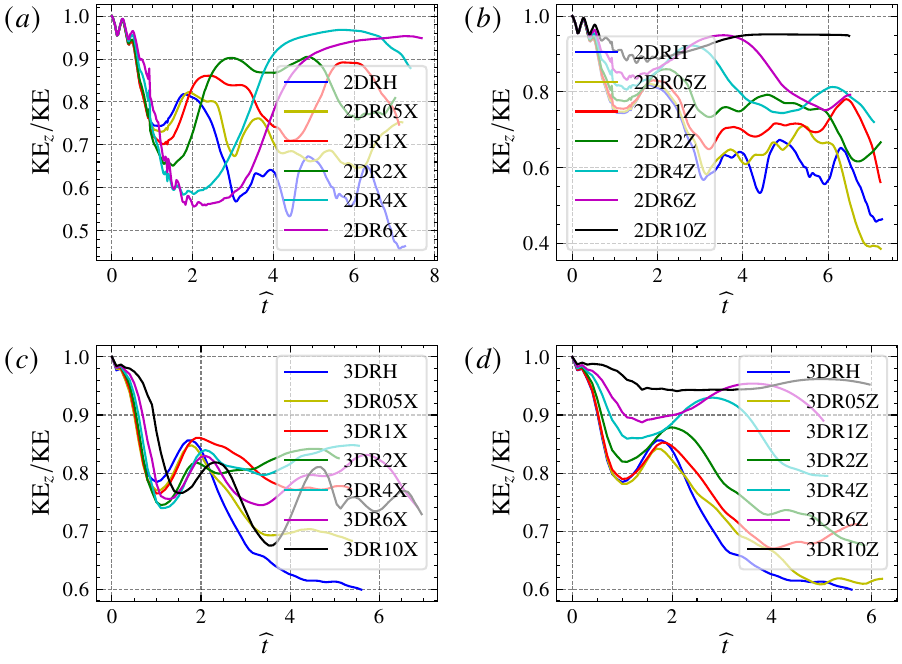}
\end{subfigure}
\renewcommand{\figurename}{Figure}\\
\caption{Ratio of vertical kinetic energy to total kinetic energy for 2‑D cases (a,b) and 3‑D cases (c,d), shown for horizontal (a,c) and vertical (b,d) magnetic fields.}
\label{fig:2D3DKEzratio}
\end{figure*}

To quantify the anisotropy, Fig.~\ref{fig:2D3DKEzratio} presents the ratio of vertical KE to total KE for hRTI and mRTI cases. For parallel external magnetic fields, this ratio initially increases with field strength in both 2‑D and 3‑D, then saturates at high field strengths. In the weak‐field regime, $T_z$ ($z$-component magnetic tension) exerts little influence on bubble and spike growth. However, $T_x$ (horizontal magnetic tension) introduces flow anisotropy that amplifies vertical motion, thereby increasing the mixing‐zone width. Under strong external parallel field strengths, vertical velocity is strongly constrained by $T_z$, and the ratio of $\mathrm{KE}_z/\mathrm{KE}$ saturates, leading to a corresponding decrease in mixing‐zone width.

For vertical external magnetic fields, the $\mathrm{KE}_z / \mathrm{KE}$ ratio increases monotonically with field strength. In this configuration, $T_z$ is a second‑order effect arising from fluctuating magnetic fields and therefore does not strongly suppress vertical motion. Instead, vertical motion is progressively enhanced as the field strengthens, while horizontal motion is reduced, resulting in a continuous expansion of the mixing zone under vertical‑field conditions. 

In summary, the Lorentz force plays a central role in determining the growth and anisotropy of the mixing zone in mRTI, acting primarily through the magnetic‐tension term. Vertical magnetic tension generally stabilizes the flow, whereas horizontal magnetic tension alters the flow anisotropy and, depending on the orientation and relative strength of the external field, can either suppress or enhance instability growth.

\subsubsection{Effect on buoyancy and vorticity terms}

\begin{figure*}
\centering
\begin{subfigure}{0.99\textwidth}
\includegraphics[width=6.5in]{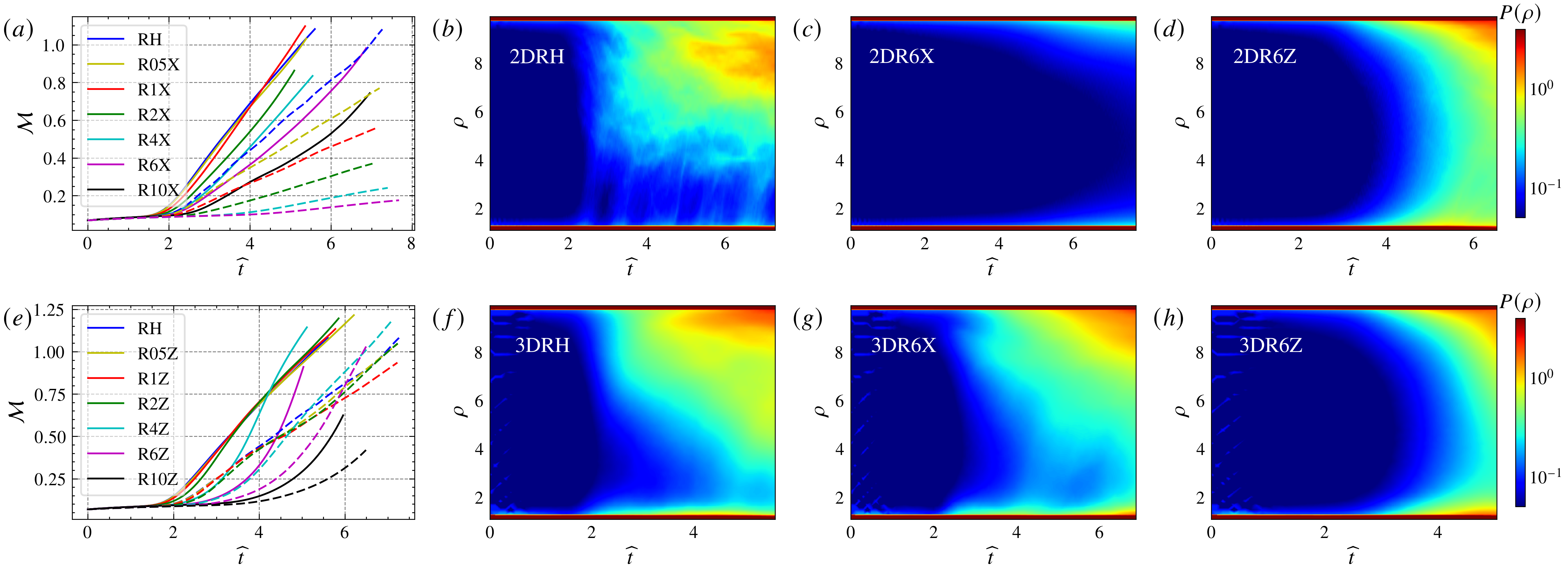}
\end{subfigure}
\renewcommand{\figurename}{Figure}\\
\caption{Mixed mass (a,e) and density‐field PDFs (b–d,f–h) over the entire simulation. Panels (a,e) compare horizontal and vertical magnetic fields, with solid lines for 3D and dashed lines for 2D. Panels (b–d,f–h) show contour plots obtained by concatenating instantaneous PDFs for hRTI, R6X, and R6Z in both 2D and 3D.}
\label{fig:mixed_rhoPDFs}
\end{figure*}

The buoyancy term is influenced by the magnetic field primarily through its impact on the density difference, or effective Atwood number. During RTI evolution, heavy and light fluids mix near the interface, reducing the density contrast and thus lowering the effective Atwood number, which in turn diminishes buoyancy. Such mixing hinders the vertical motion of both bubbles and spikes.  

Figure~\ref{fig:mixed_rhoPDFs} quantifies this effect via the mixed mass, $\mathcal{M} = \int 4\rho\,Y(1-Y)\,dV$ \cite{Zhou16PoP}, shown in panels (a) and (e), and the time‐resolved density PDFs in panels (b–d) and (f–h). The mixed mass is a conserved quantity that measures the amount of heavy fluid entrained into the light fluid. For parallel external magnetic fields (Fig.~\ref{fig:mixed_rhoPDFs}(a)), $\mathcal{M}$ decreases monotonically with increasing field strength in both 2D (dashed) and 3D (solid) cases. For vertical external fields (Fig.~\ref{fig:mixed_rhoPDFs}(e)), when $B_0 \le 0.4B_c$ only small fluctuations in $\mathcal{M}$ are observed, whereas stronger fields lead to significant suppression of mixing.  Thus, stronger external magnetic fields generally result in reduced mixing.  In addition, the normalized mixed mass, $\Psi \equiv \frac{\int \rho\, Y(1-Y)\, dV}{\int \langle \rho \rangle \langle Y \rangle \langle 1-Y \rangle\, dV}$, shown in Figure~\ref{Appfig:2D3D_mix_broad} in the Appendix, further illustrates the suppression of molecular mixing by both horizontal and vertical external magnetic fields in both 2D and 3D configurations.

The temporal evolution of the density PDFs demonstrates how magnetic fields alter the mixing process. Initially, the PDFs are close to delta functions at $\rho_l$ and $\rho_h$; as mixing develops, intermediate densities emerge. In hRTI [panels (b) and (f)], the PDF peaks shift toward intermediate densities at late times, with more efficient mixing on the heavy‐fluid side due to the $\rho$-weighted diffusivity term $\nabla\cdot(\rho D\nabla Y)$. In the presence of magnetic fields, late‐time mixing is less pronounced, and the PDFs remain more concentrated near the pure‐fluid densities $\rho_h$ and $\rho_l$. The 2D parallel‐field case 2DR6X shows the least mixing, as its PDF changes very little over time.  This is because in 2D mRTI with a horizontal magnetic field, a moderate-to-strong field suppresses the only available transverse degree of freedom, so vertical bubble–spike growth is strongly arrested. Magnetic tension can smooth or partially restore the density interface, leading to little mixing during the flow evolution.

\begin{figure*}
\centering
\begin{subfigure}{0.99\textwidth}
\includegraphics[width=5.in]{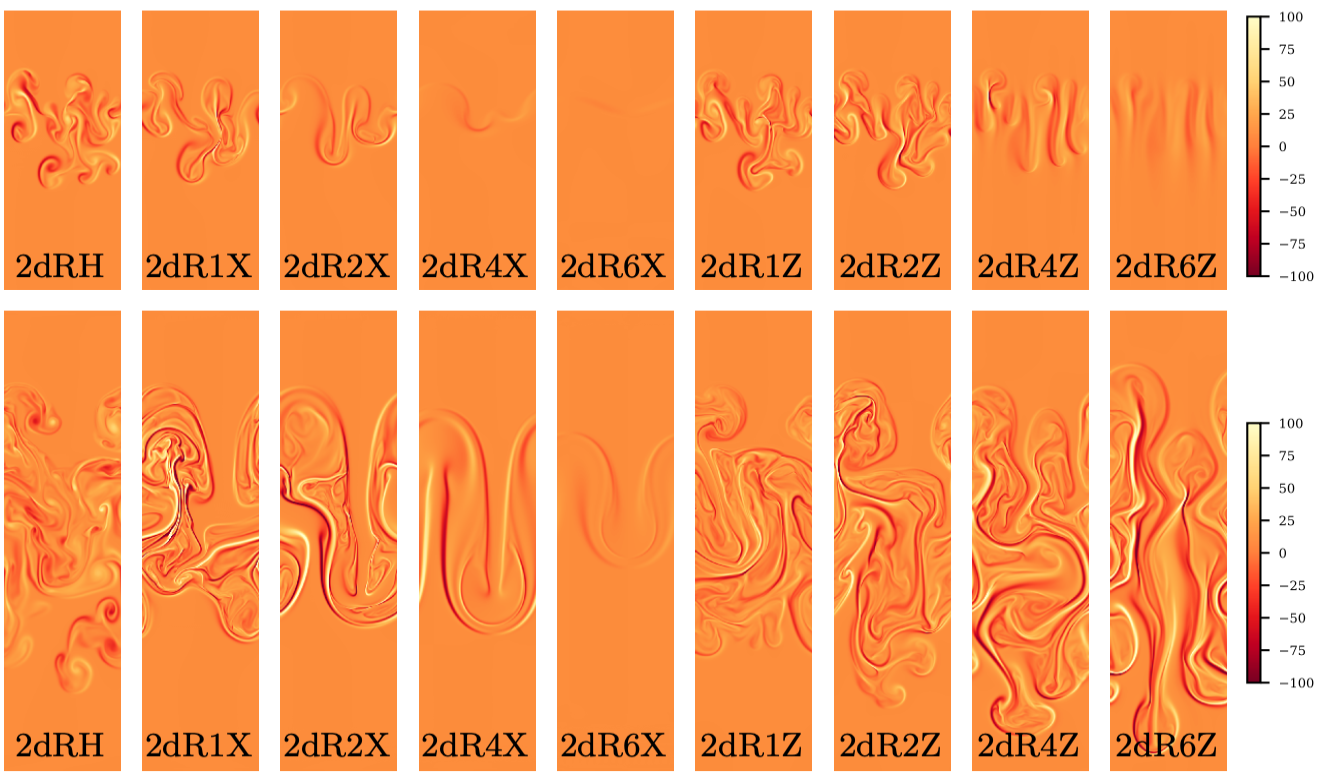}
\end{subfigure}
\renewcommand{\figurename}{Figure}\\
\caption{Visualization of dimensionless vorticty $\omega^* = (\partial_z u_x - \partial_x u_z)/\sqrt{Ag/\lambda}$ at $Agt^2/L = 10$ (top panels) and 30 (bottom panels) of 2D simulations. Note we crop the plots vertically to save space.
}
\label{fig:2DVortViz}
\end{figure*}

\begin{figure*}
\centering
\begin{subfigure}{0.99\textwidth}
\includegraphics[width=5. in]{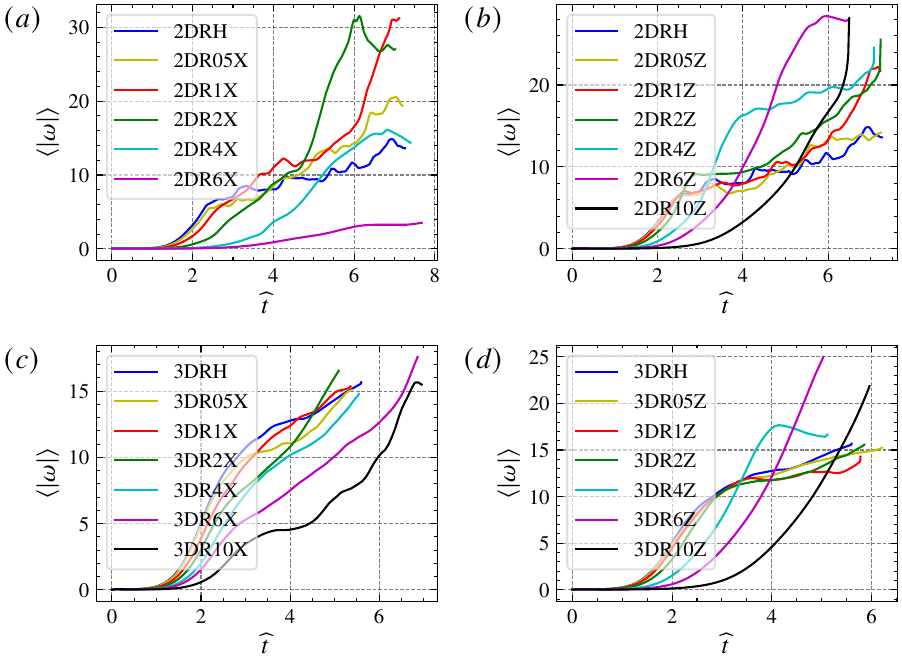}
\end{subfigure}
\renewcommand{\figurename}{Figure}\\
\caption{Mean vorticity magnitude for 2D cases in panels (a,b), and for 3D cases in panels (c,d). The strain-rate is normalized by $\sqrt{Ag/L}$.
}
\label{fig:2D3DVort}
\end{figure*}

In addition to buoyancy, external magnetic fields significantly alter the vorticity field. In classical RTI \cite{xbian2019singlemode,luo2020effects} and ablative RTI \cite{yan2016three,zhang2018self,zhang2018nonlinear}, vortices form around the spike interface as it penetrates into the lighter fluid and are advected toward the bubble tip, where they generate a centrifugal force that propels the bubble front forward. The 2‑D hRTI simulation in Fig.~\ref{fig:2DVortViz} (case 2DRH) reveals the complex vortical motions associated with this process. In mRTI, the presence of a magnetic field confines vortex generation and transport primarily to regions close to the interface. For weak horizontal fields, the peak vorticity slightly exceeds that of the hydrodynamic case, and vortex patterns, while less spatially intricate, appear more intense, as illustrated in cases 2DR1X and 2DR2X. At stronger horizontal field strengths (cases 2DR4X and 2DR6X), vortices within the bubbles are almost completely eliminated. Vertical magnetic fields, on the other hand, enhance vortex generation and transport over the full range of field strengths studied, producing stronger and more persistent vortex structures than in hRTI. These qualitative observations are quantitatively supported by the mean‐vorticity evolutions shown in Fig.~\ref{fig:2D3DVort}, which reveal enhanced vorticity production at late time in all vertical‐field runs and in weak horizontal‐field cases, but pronounced suppression in strong horizontal fields.

\subsubsection{Effect on the drag term}

\begin{figure*}
\centering
\begin{subfigure}{0.99\textwidth}
\includegraphics[width=5. in]{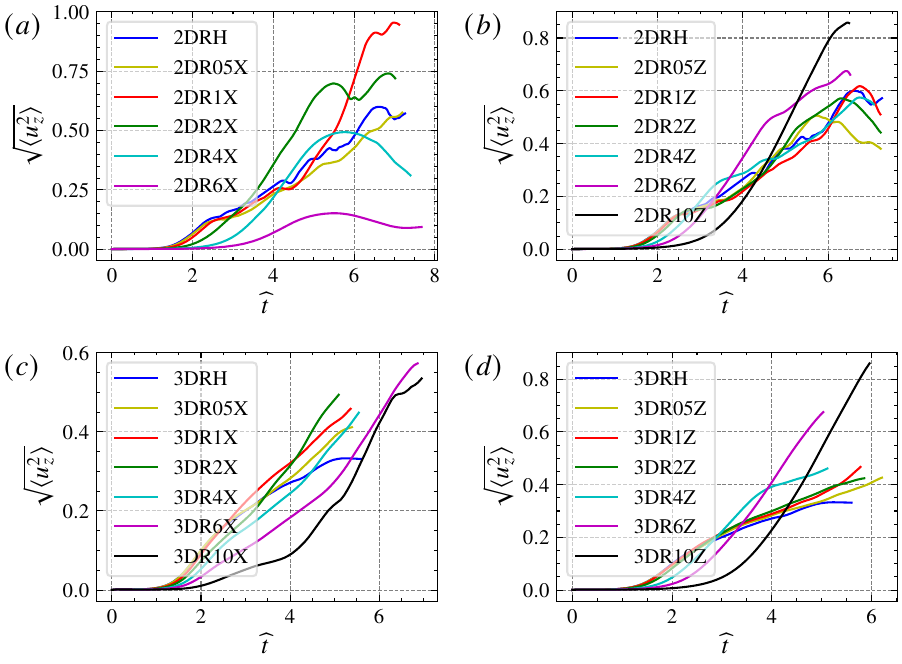}
\end{subfigure}
\renewcommand{\figurename}{Figure}\\
\caption{The mean vertical velocity for 2D in panels (a,b), and for 3D in panels (c,d). The velocity is normalized by $\sqrt{AgL}$.
}
\label{fig:2D3Duz}
\end{figure*}

\begin{figure*}
\centering
\begin{subfigure}{0.99\textwidth}
\includegraphics[width=5. in]{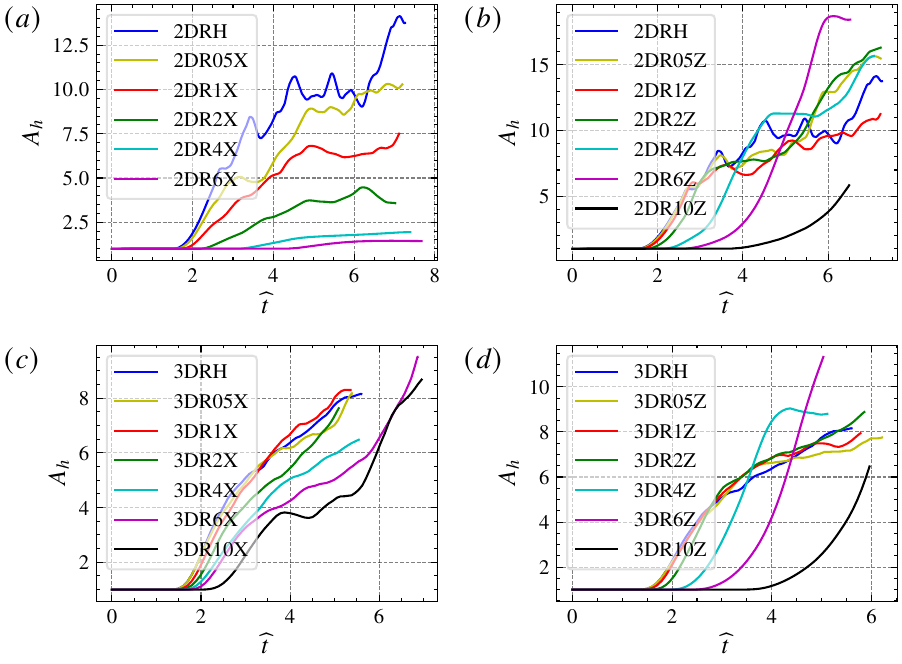}
\end{subfigure}
\renewcommand{\figurename}{Figure}\\
\caption{Total interface area projected onto the horizontal plane, shown for 2‑D cases in panels (a) and (b), and for 3‑D cases in panels (c) and (d). All areas are normalized by $L^2$.
}
\label{fig:2D3Darea}
\end{figure*}

We now investigate the influence of external magnetic fields on the drag term in Eq.~\eqref{eq:bub_dynamics}. This term scales with combinations of the vertical bubble velocity and the area projected onto the horizontal plane. The time evolution of the root–mean–square vertical velocity, $\sqrt{\langle u_z^2\rangle}$, is presented in Fig.~\ref{fig:2D3Duz}, while Fig.~\ref{fig:2D3Darea} shows the corresponding projected total interface area. Since vertical velocity is approximately the time derivative of the mixing–zone width, it is expected that the two quantities follow the same trends, as is confirmed in Fig.~\ref{fig:2D3Duz}. The evolution of interfacial area $A_h$ (horizontal projected area; evolutions of total area are similar) is slightly different: for parallel external fields in 2‑D, the interface area decreases steadily as the magnetic field becomes stronger; in 3‑D, this quantity remains comparable across weak‐field cases, but show a marked decrease when the field is strong. In contrast, vertical‐field configurations in both 2‑D and 3‑D display similar interface areas at weak field strength, while at strong field strength the growth is initially delayed and then subsequently increases, which corresponds to a delayed but eventual enhancement trends as in the mixing-zone growth. These behaviors indicate that magnetic tension can reduce drag by suppressing bubble rise speeds and limiting projected area, with the strongest suppression occurring under strong parallel fields.

Overall, magnetic fields regulate bubble dynamics in mRTI through coupled modifications of flow morphology, anisotropy, vertical velocity, and vorticity generation. In parallel external fields, weak field strengths alter the flow anisotropy to favor vertical motion, sustaining or even enhancing vorticity near the interface. The external field also reduces interfacial mixing, thereby  enhancing buoyancy and accelerating the growth of the mixing width. In contrast, strong parallel fields suppress instability growth by limiting vertical motion through magnetic tension and diminishing vorticity production in the vicinity of bubble and spike tips.  
Under vertical external fields, the interplay of preferred vertical anisotropy, intensified vorticity generation, and reduced molecular‐level mixing produces a delayed but ultimately amplified influence on instability growth as field strength increases. These interconnected effects account for the distinctly different nonlinear evolution pathways observed across parallel- and vertical-field mRTI configurations.

\section{Summary}
This study examined the influence of parallel (horizontal) and vertical external magnetic fields on the nonlinear evolution of mRTI using 2-D and 3-D resistive MHD simulations, with comparisons to hydrodynamic RTI cases. By varying both the strength and orientation of the applied fields, we quantified their roles in shaping mixing‐zone growth, flow anisotropy, vorticity generation, and interfacial areas.  

For parallel external magnetic fields, weak field strengths were found to enhance mixing‐zone growth by modifying the flow anisotropy to favor vertical motion, while the buoyancy term and sustained vorticity near the fluid interface are little affected. Under these conditions, the ratio of vertical to total kinetic energy increases with field strength. At stronger horizontal field strengths, magnetic tension limited vertical motion, the kinetic energy ratio saturated, vorticity near bubble and spike tips was reduced, and the overall mixing‐zone width decreased. In contrast, vertical magnetic fields initially delayed mixing‐zone evolution but later, once the system entered self‐similar growth, promoted faster expansion of the mixing zone through enhanced vertical anisotropy and increased vorticity. In these cases, the vertical‐to‐total kinetic energy ratio increased steadily with field strength, while the vertical magnetic tension in this geometry is not strong enough to oppose bubble and spike motions.  

The physical picture that emerges is that magnetic fields influence mRTI primarily through the Lorentz‐force magnetic tension term by modifying anisotropy, while buoyancy remains the fundamental driver of instability. In addition, by altering the degree of mixing, the magnetic field modifies the effective Atwood number and thus the buoyancy force. The drag force, proportional to the product of vertical velocity and interfacial area, and the vorticity field, which accelerates bubbles and spikes via centrifugal forces, are also modulated in ways that depend on both field orientation and strength. Parallel fields tend to suppress the overall vorticity magnitude but can locally enhance it near the interface in the weak‐field regime, whereas vertical fields initially suppress and later enhance vorticity growth.  

These trends are broadly consistent with earlier studies. Our results match the findings of Carlyle and Hillier \cite{carlyle2017non}, who reported suppression of 3D mRTI growth under strong horizontal fields, and of Kalluri and Hillier \cite{Kalluri2025JFM}, who observed increased mixing width for weak horizontal fields. They also agree with recent reports on vertical fields under Boussinesq description showing delayed but ultimately enhanced growth \cite{Briard2024JFM,SinghPal2025JFM}. Differences from the results of Stone and Gardiner \cite{stone2007nonlinear}, who found that moderately strong horizontal fields produced greater mixing than the hydrodynamic case, likely stem from their ideal‐MHD treatment without viscosity or magnetic diffusivity, whereas these effects are explicitly included in our simulations. Analytical considerations further support the interpretation that in the limit of very strong parallel magnetic fields, bounded drag, buoyancy, and vorticity effects are overwhelmed by the unbounded vertical Lorentz force term, leading inevitably to suppression of mRTI relative to hydrodynamic RTI.  

In conclusion, magnetic field orientation plays a decisive role in determining mRTI evolution. Parallel fields can either enhance or suppress instability growth depending on their strength, while vertical fields tend to delay growth at early times but ultimately extend the mixing zone at late times. These findings provide a coherent framework linking anisotropy, Lorentz forces, drag, buoyancy, and vorticity to the nonlinear dynamics of mRTI. Looking forward, future work will incorporate additional physics, including ablation effects, Biermann‐battery‐generated magnetic fields, spatially varying resistivity, and their coupled impacts on instability growth. Three‐dimensional simulations in cylindrical geometry will also be pursued to assess axial‐field effects relevant to inertial confinement fusion, where geometric differences may lead to qualitatively different behavior from the Cartesian results presented here.

\section*{Acknowledgments}
This research was funded by US NSF grant PHY-2206380. 
HA was also supported by US DOE grants DE-SC0020229, DE-SC0014318, DE-SC0019329, US NSF grants PHY-2020249, OCE-2123496, OCE-2446475, TI-2449340, and by U.S. NNSA grants DE-NA0004144, DE-NA0003914, and DE-NA0004134.
Computing time was provided by NERSC under Contract No. DE-AC02-05CH11231, and the Texas Advanced Computing Center (TACC) at The University of Texas at Austin, under ACCESS allocation grant EES220052.

\section*{DATA AVAILABILITY}
The data that supports the findings of this study are available from the corresponding author upon reasonable request.

\section{Appendix}

\subsection{Simulation parameters}
All simulation cases are summarized in Table~\ref{Tbl:Simulations}, which lists the 2‑D and 3‑D runs performed under various horizontal and vertical external magnetic fields. Each case was repeated for three magnetic Prandtl numbers, $\mathrm{Pr}_m = 1$, $0.2$, and $5$. The results presented in the main text correspond to $\mathrm{Pr}_m = 1$, while the outcomes for the other two values are provided in Section~\ref{appendix:nonunity_Prm}.

\begin{table*}[]
\caption{Simulation parameters. Resolution is $N_x \times N_z$ in 2D and  $N_x \times N_y\times N_z$ in 3D. Domain size is $L_x \times L_z$ in 2D and  $L_x \times L_y\times L_z$ in 3D. Kinematic viscosity $\nu \equiv 2\mu / (\rho_h + \rho_l)$. Each of the Run shown in the table is repeated for magnetic Prandtl number $\mathrm{Pr}_m = \nu/\eta =0.2$ and 5, where $\eta=4.55\times 10^{-6}$ and $\eta=1.82\times 10^{-7}$, respectively.}
\begin{center}
\begin{tabular}{lccccccccc}
\hline
\hline
    Run          & $\bB_0$ direction & $|\bB_0|$ & Dimensions  &$\nu$& $\eta$   & Resolution               & Domain size                    \\ \hline
    2DRH         & -              & -            & 2D          & $9.10\times 10^{-7}$  & -       & $512\times2048$           & $0.1\times0.4$             \\
    2DR05X        & $x$            & $0.05B_c$     & 2D          & $9.10\times 10^{-7}$  & $9.10\times 10^{-7}$       & $512\times2048$           & $0.1\times0.4$        \\
    2DR1X        & $x$            & $0.1B_c$     & 2D          & $9.10\times 10^{-7}$  & $9.10\times 10^{-7}$       & $512\times2048$           & $0.1\times0.4$        \\
    2DR2X        & $x$            & $0.2B_c$     & 2D          & $9.10\times 10^{-7}$  & $9.10\times 10^{-7}$       & $512\times2048$           & $0.1\times0.4$      \\
    2DR4X        & $x$            & $0.4B_c$     & 2D          & $9.10\times 10^{-7}$  & $9.10\times 10^{-7}$        & $512\times2048$           & $0.1\times0.4$       \\ 
    2DR6X        & $x$            & $0.6B_c$     & 2D          & $9.10\times 10^{-7}$  & $9.10\times 10^{-7}$       & $512\times2048$           & $0.1\times0.4$      \\ 
    2DR05Z        & $z$            & $0.05B_c$     & 2D          & $9.10\times 10^{-7}$  & $9.10\times 10^{-7}$       & $512\times4096$           & $0.1\times0.8$      \\
    2DR1Z        & $z$            & $0.1B_c$     & 2D          & $9.10\times 10^{-7}$  & $9.10\times 10^{-7}$       & $512\times4096$           & $0.1\times0.8$      \\
    2DR2Z        & $z$            & $0.2B_c$     & 2D          & $9.10\times 10^{-7}$  & $9.10\times 10^{-7}$       & $512\times4096$           & $0.1\times0.8$       \\
    2DR4Z        & $z$            & $0.4B_c$     & 2D          & $9.10\times 10^{-7}$  & $9.10\times 10^{-7}$       & $512\times4096$           & $0.1\times0.8$       \\ 
    2DR6Z        & $z$            & $0.6B_c$     & 2D          & $9.10\times 10^{-7}$  & $9.10\times 10^{-7}$       & $512\times4096$           & $0.1\times0.8$       \\ 
    2DR10Z       & $z$            & $1.0 B_c$       & 2D          & $9.10\times 10^{-7}$  & $9.10\times 10^{-7}$       & $512\times4096$           & $0.1\times0.8$       \\ 
    3DRH         &  -             & -            & 3D          & $1.36\times 10^{-6}$  & -             & $256\times256\times1024$  & $0.1\times0.1\times0.4$         \\
    3DR05X        & $x$            & $0.05B_c$     & 3D          & $1.36\times 10^{-6}$  & $1.36\times 10^{-6}$       & $256\times256\times1024$  & $0.1\times0.1\times0.4$        \\
    3DR1X        & $x$            & $0.1B_c$     & 3D          & $1.36\times 10^{-6}$  & $1.36\times 10^{-6}$       & $256\times256\times1024$  & $0.1\times0.1\times0.4$        \\
    3DR2X        & $x$            & $0.2B_c$     & 3D          & $1.36\times 10^{-6}$  & $1.36\times 10^{-6}$       & $256\times256\times1024$  & $0.1\times0.1\times0.4$        \\
    3DR4X        & $x$            & $0.4B_c$     & 3D          & $1.36\times 10^{-6}$  & $1.36\times 10^{-6}$       & $256\times256\times1024$  & $0.1\times0.1\times0.4$      \\
    3DR6X        & $x$            & $0.6B_c$     & 3D          & $1.36\times 10^{-6}$  & $1.36\times 10^{-6}$       & $256\times256\times1024$  & $0.1\times0.1\times0.4$        \\
    3DR10X       & $x$            & $1.0B_c$     & 3D          & $1.36\times 10^{-6}$  & $1.36\times 10^{-6}$       & $256\times256\times1024$  & $0.1\times0.1\times0.4$        \\ 
    3DR05Z        & $z$            & $0.05B_c$     & 3D          & $1.36\times 10^{-6}$  & $1.36\times 10^{-6}$        & $256\times256\times1024$  & $0.1\times0.1\times0.4$       \\
    3DR1Z        & $z$            & $0.1B_c$     & 3D          & $1.36\times 10^{-6}$  & $1.36\times 10^{-6}$        & $256\times256\times1024$  & $0.1\times0.1\times0.4$       \\
    3DR2Z        & $z$            & $0.2B_c$     & 3D          & $1.36\times 10^{-6}$  & $1.36\times 10^{-6}$        & $256\times256\times1024$  & $0.1\times0.1\times0.4$        \\
    3DR4Z        & $z$            & $0.4B_c$     & 3D          & $1.36\times 10^{-6}$  & $1.36\times 10^{-6}$        & $256\times256\times1024$  & $0.1\times0.1\times0.4$      \\ 
    3DR6Z        & $z$            & $0.6B_c$     & 3D          & $1.36\times 10^{-6}$  & $1.36\times 10^{-6}$        & $256\times256\times1024$  & $0.1\times0.1\times0.4$        \\ 
    3DR10Z       & $z$            & $1.0B_c$     & 3D          & $1.36\times 10^{-6}$  & $1.36\times 10^{-6}$        & $256\times256\times1024$  & $0.1\times0.1\times0.4$        \\ 
\hline
\hline
\end{tabular}
\end{center}
\label{Tbl:Simulations}
\end{table*}

\subsection{Linear stability analysis}

\begin{figure}
\centering
\begin{subfigure}{0.4\textwidth}
\includegraphics[width=\textwidth]{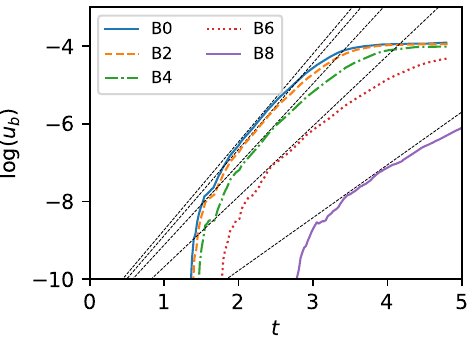}
\end{subfigure}
\caption{Growth rates for single‑mode simulations with a uniform horizontal magnetic field parallel to the interface. Labels B0–B8 indicate the magnetic‑field strength, varying from $0.0\,B_c$ to $0.8\,B_c$. The dashed line represents the prediction from linear stability analysis \cite{chandrasekhar1961hydrodynamic}.}
\label{fig:lsa}
\end{figure}

We validate our solver by comparing numerical results with predictions from linear stability analysis \cite{chandrasekhar1961hydrodynamic}. For a parallel magnetic field $\mathbf{B}$, the linear growth rate is given by Eq.~\eqref{eq:lsa}. We perform 2D single‑mode simulations with parallel magnetic‑field strengths $|\mathbf{B}| = 0$, $0.2B_c$, $0.4B_c$, $0.6B_c$, and $0.8B_c$. The computational domain is $L_x \times L_z = 0.1 \times 0.2$ with a grid resolution of $256 \times 512$. Other parameters are $\rho_h = 10$, $\rho_l = 1$, and $g = 0.1$.  

Figure~\ref{fig:lsa} compares the measured bubble velocity amplitude $u_b$ with the theoretical prediction from Eq.~\eqref{eq:lsa}. After the initial sublinear phase, the simulations closely follow the linear‑theory prediction. As expected, the growth rate decreases with increasing magnetic‑field strength. Upon entering the nonlinear stage, the growth rate saturates.  

\subsection{Comparison of Exponential and Gaussian Initial Conditions}
\label{appendix:IC_compare}

\begin{table}[]
\caption{Parameters for the initial‑condition study. The computational domain is $L_x \times L_z = 0.1 \times 0.2$, where $N_x$ and $N_z$ denote the grid resolutions in the $x$ and $z$ directions, respectively.}
\begin{center}
\begin{tabular}{lcccccc}
\hline
\hline
    ID      & $N_x$  & $N_z$     \\ \hline
    0        & 1024  & 2048     \\ 
    1        & 128   & 256      \\
    2        & 256   & 512      \\
    3        & 512   & 1024     \\
    4        & 128   & 2048     \\ 
\hline
\hline
\end{tabular}
\end{center}
\label{Tbl:ICS_sims}
\end{table}

\begin{figure*}
\centering
\begin{subfigure}{0.45\textwidth}
\includegraphics[width=2.4 in]{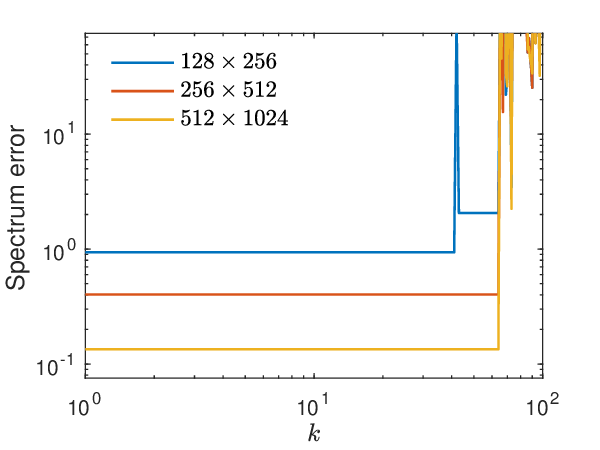}
\caption{ID1-ID3}
\end{subfigure}
\begin{subfigure}{0.45\textwidth}
\includegraphics[width=2.4 in]{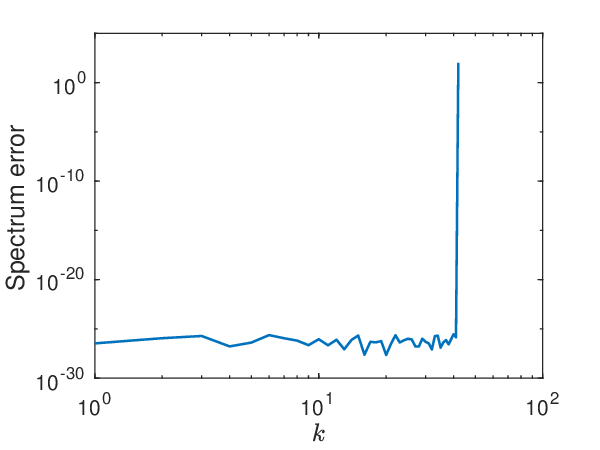}
\caption{ID4}
\end{subfigure}
\renewcommand{\figurename}{Figure}
\caption{Percentage error of initial velocity spectrum using IC1. (a) ID1-ID3 (b) ID4, which has the same resolution in $z$ as that the benchmark ID0. The percentage error is defined by $100\times|\hat{u}_z(k) - \hat{u}_z^0(k)|^2 / |\hat{u}_z^0(k)|^2$. IC1 decays exponentially in $z$ direction.}
\label{appendix:IC1_error}
\end{figure*}

\begin{figure*}
\centering
\begin{subfigure}{0.45\textwidth}
\includegraphics[width=2.4 in]{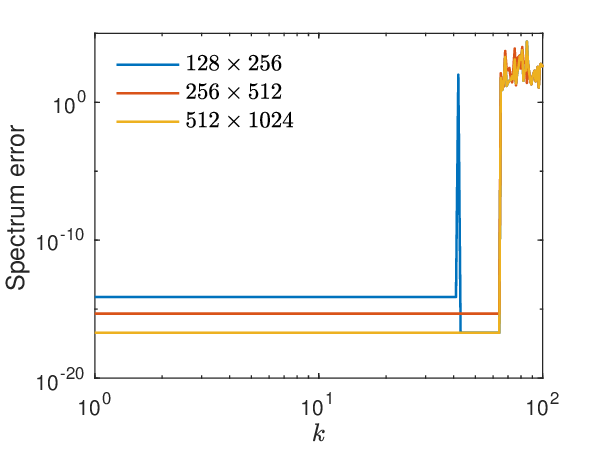}
\end{subfigure}
\renewcommand{\figurename}{Figure}
\caption{Percentage error of initial velocity spectrum using IC2. The percentage error is defined by $100\times|\hat{u}_z(k) - \hat{u}_z^0(k)|^2 / |\hat{u}_z^0(k)|^2$. IC2 decays as Gaussain in $z$ direction. }
\label{appendix:IC2_error}
\end{figure*}

We compare two types of initial conditions (ICs). The initial perturbation is given by  
$$v_p(x,z) = \sum_{m} v_{pk} \cos(m k_l x + \phi_{k0})$$  
where $k_l = 2\pi / L_x = 2\pi / L_y$ and $\phi_{k0}$ is a random phase. The perturbation amplitude decays with $z$ according to either  
$$v_{pk} = v_{pk0} \exp(-k_l |z - z_0|) \quad \text{(IC1, exponential decay)}$$
or  
$$v_{pk} = v_{pk0} \exp\left(-\frac{k_l^2 |z - z_0|^2}{\pi^2}\right) \quad \text{(IC2, Gaussian decay)}$$   
where $z_0$ is the initial interface location. For simplicity, $v_{pk0}$ is constant for all modes, with mode numbers ranging from 1 to 64. The only difference between IC1 and IC2 lies in the decay function along $z$.

We evaluate the quality of each IC by computing the initial error spectrum at the perturbation plane. The percentage error is defined as:  
$\text{Error}(k) = 100 \times |\hat{u}_z(k) - \hat{u}_z^0(k)|^2/|\hat{u}_z^0(k)|^2$,  
where $\hat{u}_z(k)$ is the Fourier transform of the initial vertical velocity and $\hat{u}_z^0(k)$ corresponds to the highest resolution case ($N_x \times N_z = 1024 \times 2048$), which is taken as the benchmark. Since the perturbation is applied only to the vertical velocity, we restrict our analysis to $\hat{u}_z(k)$. Error spectra are computed at various grid resolutions, as is listed in Table~\ref{Tbl:ICS_sims}; the highest‑resolution result $\hat{u}_z^0$ (ID0 in Table~\ref{Tbl:ICS_sims}) is used for reference.   

Results for IC1 (Fig.~\ref{appendix:IC1_error}) show a roughly constant error across all modes for a given resolution. The error increases from approximately $0.09\%$ to $1\%$ when the grid resolution is reduced from $512 \times 1024$ to $128 \times 256$ (Fig.~\ref{appendix:IC1_error}(a)). In contrast, for a resolution of $128 \times 2048$, the error is negligible (Fig.~\ref{appendix:IC1_error}(b)), suggesting that the primary source of error lies in the $z$-direction, arising from the exponential decay profile and/or discretization schemes.  
When IC2 (Gaussian decay) is used, the error spectrum (Fig.~\ref{appendix:IC2_error}) is smaller than $10^{-20}\%$, indicating a dramatic reduction in initial discretization error compared with IC1. Based on these results, the Gaussian‑profile initial condition (IC2) is adopted for this study.  

\subsection{Convergence study}\label{appendix:convergence_study}

\begin{figure}
\begin{subfigure}{0.4\textwidth}
\includegraphics[width=\textwidth]{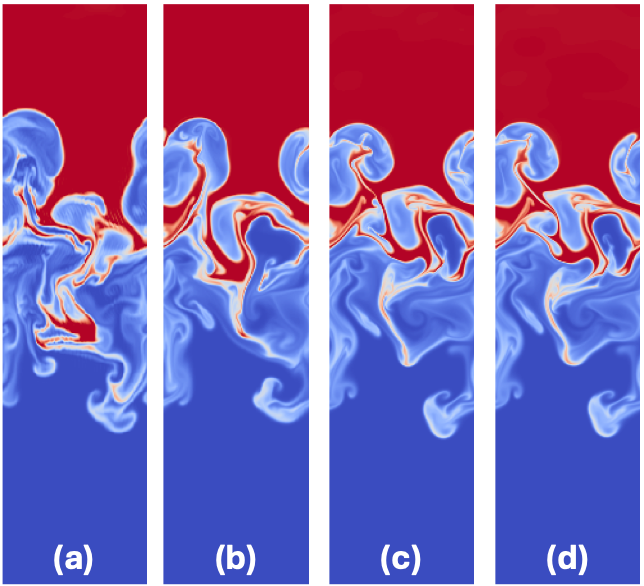}
\end{subfigure}
\renewcommand{\figurename}{Figure}
\caption{Density field $\rho$ from 2D multi‑mode hRTI simulations with $\mu = 5 \times 10^{-6}$ at $A g t^2 / L \approx 23$. Grid resolutions: (a) $N_x \times N_z = 128 \times 512$, (b) $256 \times 1024$, (c) $512 \times 2048$, and (d) $1024 \times 4096$.}
\label{appendix:2d_conver_visual}
\end{figure}

\begin{figure*}
\centering
\begin{subfigure}{0.45\textwidth}
\includegraphics[width=2.5 in]{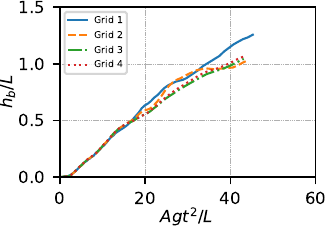}
\end{subfigure}
%\mbox{\hspace{0.5cm}}
\begin{subfigure}{0.45\textwidth}
\includegraphics[width=2.5 in]{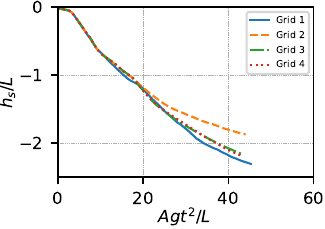}
\end{subfigure}
\renewcommand{\figurename}{Figure}\\
\caption{Time evolution of bubble ($h_b$) and spike ($h_s$) heights in the 2D convergence study with $\mu = 5\times 10^{-6}$. Grid 1-4 resolutions are: $128 \times 512$, $256 \times 1024$, $512 \times 2048$, and $1024 \times 4096$.}
\label{appendix:2dConverge}
\end{figure*}

In this subsection, we present the convergence study. Figure~\ref{appendix:2d_conver_visual} shows density fields at various grid resolutions for $\mu = 5\times 10^{-6}$. The flow structures are nearly identical at $N_x \times N_z = 512 \times 2048$ and $1024 \times 4096$, indicating that the solution converges at the $512 \times 2048$ resolution. Furthermore, Figure~\ref{appendix:2dConverge} demonstrates that the time evolution of bubble and spike heights obtained with the $512 \times 2048$ grid agrees well with that from the $1024 \times 4096$ grid, which is the primary quantity of interest in this study. The convergence path from coarse to fine resolution is not strictly monotonic for bubble and spike heights, underscoring the importance of a systematic convergence analysis when comparing mixing‑zone widths across different mRTI cases.  

Based on these results, we select the physical viscosity and grid resolution to ensure the smallest viscosity compatible with the chosen resolution. For 2D simulations, we use $\mu = 5\times 10^{-6}$ with a $512 \times 2048$ grid (or $512 \times 4096$ for an aspect ratio of 8). For 3D mRTI simulations, we use $\mu = 7.5\times 10^{-6}$ on a $256 \times 256 \times 1024$ grid.

\subsection{The normalized mixed mass}

\begin{figure*}
\centering
\begin{subfigure}{0.99\textwidth}
\includegraphics[width=5 in]{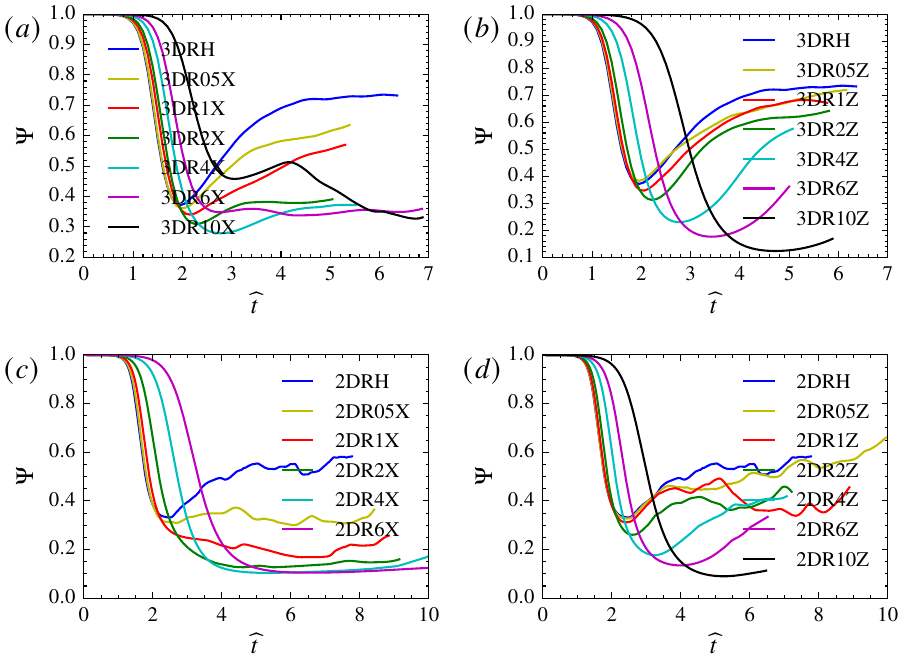}
\end{subfigure}
\renewcommand{\figurename}{Figure}\\
\caption{Temporal evolution of the normalized mixed mass ($\Psi$) for (a) 3D horizontal mRTI, (b) 3D vertical mRTI, (c) 2D horizontal mRTI, and (d) 2D vertical mRTI.}
\label{Appfig:2D3D_mix_broad}
\end{figure*}

Figure~\ref{Appfig:2D3D_mix_broad} shows the normalized mixed mass, $\Psi$, for 2D and 3D mRTI cases subjected to different external magnetic field orientations. The normalized mixed mass \citep{Zhou16PoP,Zhou2024Book} quantifies the degree of molecular mixing in RT flows and is defined as
$\Psi = \frac{\int \rho\, Y(1-Y) \, dV}{\int \langle \rho \rangle \langle Y \rangle \langle 1-Y \rangle \, dV}$,
where $\langle \cdot \rangle$ denotes averaging over horizontal homogeneous directions.  

For hydrodynamic RTI in 3D, $\Psi$ asymptotes to a constant value of approximately 0.7, whereas in 2D the asymptotic value is lower, indicating reduced mixing. In magnetic RTI cases, both horizontal (panels a,c) and vertical (panels b,d) magnetic fields induce an almost monotonic reduction in $\Psi$, highlighting the suppression of molecular mixing due to magnetic tension effects.

% \begin{figure}%[hbt!]
% \centering
% \begin{minipage}[b]{1.0\textwidth}
% \centering
%     \begin{subfigure}{0.95\textwidth}  
%     \centering
%     \includegraphics[width=0.99\textwidth]{added/2D_3D_mix.pdf} 
%     \end{subfigure} \hspace{1em}
% \end{minipage}
%     \caption{Temporal evolution of the normalized mixed mass ($\Psi$) for (a) 3D horizontal mRTI, (b) 3D vertical mRTI, (c) 2D horizontal mRTI, and (d) 2D vertical mRTI.  \label{Appfig:2D3D_mix_broad}}
% \end{figure}

\subsection{mRTI simulations with varying magnetic Prandtl number} \label{appendix:nonunity_Prm}
\begin{figure*}
\centering
\begin{subfigure}{0.99\textwidth}
\includegraphics[width=6 in]{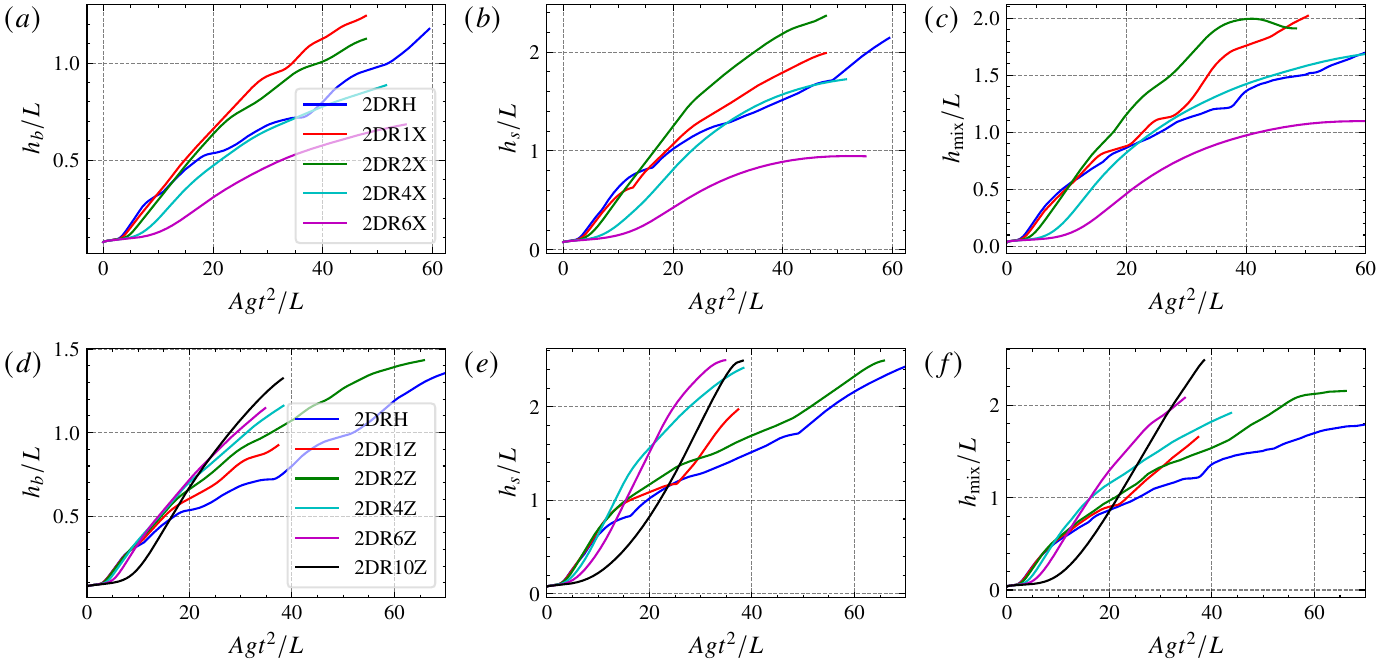}
\end{subfigure}
\renewcommand{\figurename}{Figure}\\
\caption{Time evolution of the bubble height $(a,d)$, spike height $(b,e)$, and mixing zone width $(c,f)$ for 2D simulations with $\mathrm{Pr}_m=1$. Panels (a–c) illustrate the effects of horizontal magnetic fields, whereas panels (d–f) show the effects of vertical magnetic fields.
}
\label{fig:2Dmixingheight_Pr02}
\end{figure*}

\begin{figure*}
\centering
\begin{subfigure}{0.99\textwidth}
\includegraphics[width=6 in]{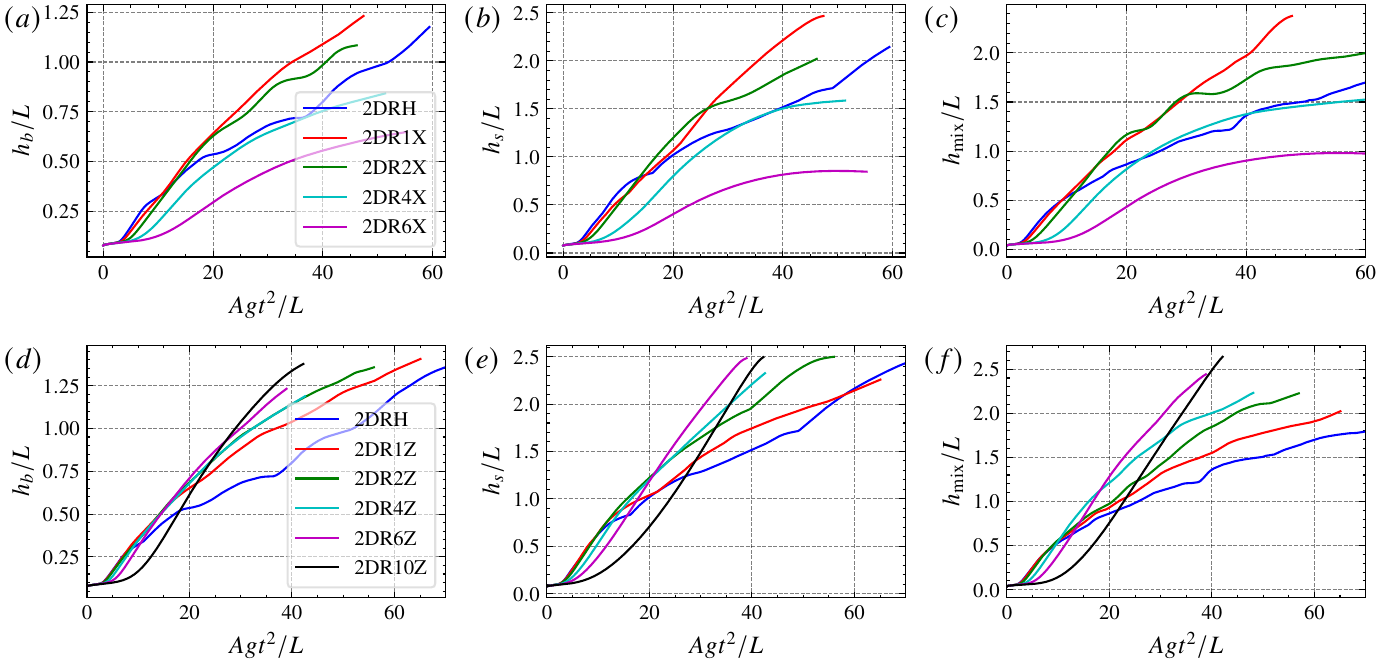}
\end{subfigure}
\renewcommand{\figurename}{Figure}\\
\caption{Time evolution of the bubble height $(a,d)$, spike height $(b,e)$, and mixing zone width $(c,f)$ for 2D simulations with $\mathrm{Pr}_m=1$. Panels (a–c) illustrate the effects of horizontal magnetic fields, whereas panels (d–f) show the effects of vertical magnetic fields.
}
\label{fig:2Dmixingheight_Pr5}
\end{figure*}

\begin{figure*}
\centering
\begin{subfigure}{0.99\textwidth}
\includegraphics[width=6 in]{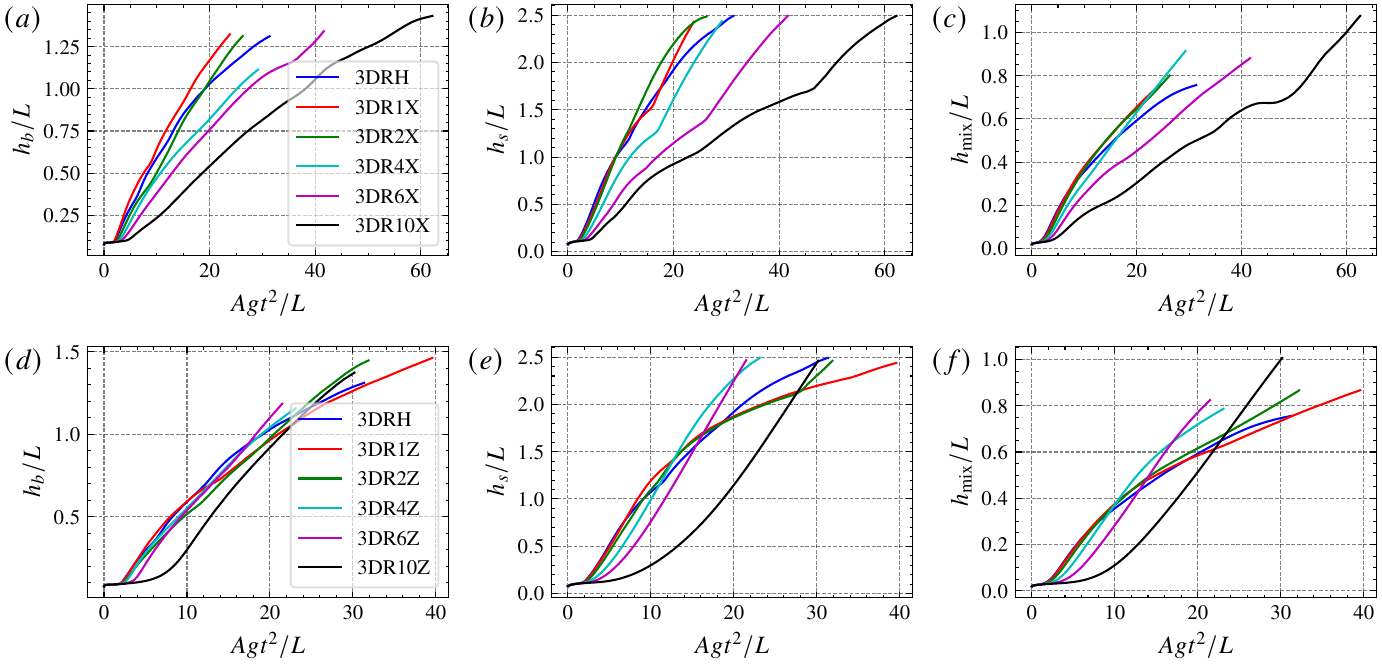}
\end{subfigure}
\renewcommand{\figurename}{Figure}\\
\caption{Time evolution of the bubble height $(a,d)$, spike height $(b,e)$, and mixing zone width $(c,f)$ for 2D simulations with $\mathrm{Pr}_m=1$. Panels (a–c) illustrate the effects of horizontal magnetic fields, whereas panels (d–f) show the effects of vertical magnetic fields.
}
\label{fig:3Dmixingheight_Pr02}
\end{figure*}

\begin{figure*}
\centering
\begin{subfigure}{0.99\textwidth}
\includegraphics[width=6 in]{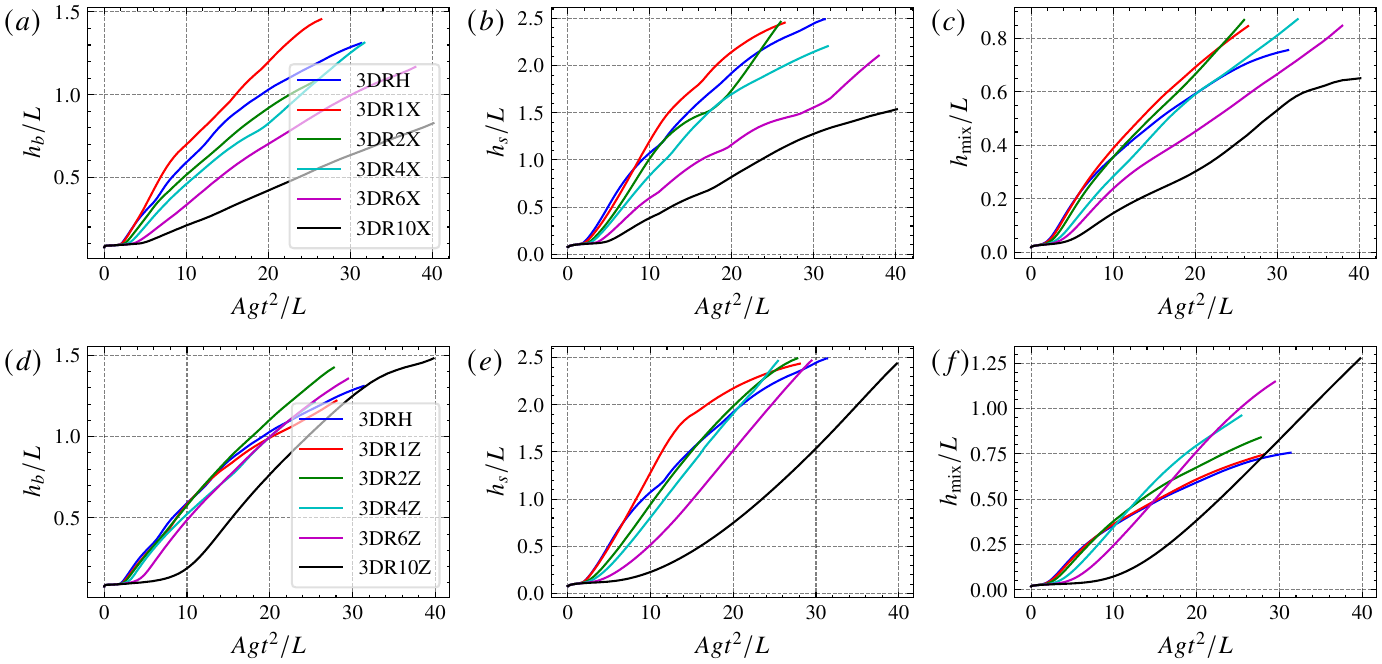}
\end{subfigure}
\renewcommand{\figurename}{Figure}\\
\caption{Time evolution of the bubble height $(a,d)$, spike height $(b,e)$, and mixing zone width $(c,f)$ for 2D simulations with $\mathrm{Pr}_m=1$. Panels (a–c) illustrate the effects of horizontal magnetic fields, whereas panels (d–f) show the effects of vertical magnetic fields.
}
\label{fig:3Dmixingheight_Pr5}
\end{figure*}

In addition to the unity magnetic Prandtl number presented in the main text, we have also included additional simulations with magnetic Prandtl numbers $\mathrm{Pr}_{m} = 0.2$ and $\mathrm{Pr}_{m} = 5$ in both 2D and 3D. The 2D results are presented in Figs.~\ref{fig:2Dmixingheight_Pr02} and \ref{fig:2Dmixingheight_Pr5} for $\mathrm{Pr}_{m} = 0.2$ and $\mathrm{Pr}_{m} = 5$, respectively, while the corresponding 3D results are shown in Figs.~\ref{fig:3Dmixingheight_Pr02} and \ref{fig:3Dmixingheight_Pr5}. In both 2D and 3D, the trends are qualitatively similar to those observed for the unity magnetic Prandtl number cases shown in Figs.~\ref{fig:2Dmixingheight} and \ref{fig:3Dmixingheight} in the main text (see also \cite{bian2021scaling}).

\bibliographystyle{abbrv}
\bibliography{main.bib}
\end{document}